\documentstyle[emulateapj]{article}

\def\eq#1{\begin{equation} #1 \end{equation}}

\def\mic              {\hbox{$\mu{\rm m}$}}
\def\about            {\hbox{$\sim$}}

\def\comm#1           {{\tt (COMMENT: #1)}}
\def\u                {\hbox{$u^*$}}
\def\g                {\hbox{$g^*$}}
\def\r                {\hbox{$r^*$}}
\def\i                {\hbox{$i^*$}}
\def\z                {\hbox{$z^*$}}
\def\ug               {\hbox{$u^*-g^*$}}
\def\ur               {\hbox{$u^*-r^*$}}
\def\gr               {\hbox{$g^*-r^*$}}
\def\gi               {\hbox{$g^*-i^*$}}
\def\ri               {\hbox{$r^*-i^*$}}
\def\iz               {\hbox{$i^*-z^*$}}
\def\rz               {\hbox{$r^*-z^*$}}
\def\Fpeak            {\hbox{F$_{peak}$}}
\def\Fint             {\hbox{F$_{int}$}}
\def\t                {\hbox{$t$}}

\begin{document}

\title{   Optical and Radio Properties of Extragalactic Sources Observed
              by the FIRST Survey and the Sloan Digital Sky Survey   }

\author{
\v{Z}eljko Ivezi\'{c}\altaffilmark{\ref{Princeton}},
Kristen Menou\altaffilmark{\ref{Princeton}},
Gillian R. Knapp\altaffilmark{\ref{Princeton}},
Michael A. Strauss\altaffilmark{\ref{Princeton}},
Robert H. Lupton\altaffilmark{\ref{Princeton}},
Daniel E. Vanden Berk\altaffilmark{\ref{Fermilab}},
Gordon T. Richards\altaffilmark{\ref{PennState}},
Christy Tremonti\altaffilmark{\ref{Baltimore}},
Michael A. Weinstein\altaffilmark{\ref{PennState}},
Scott Anderson\altaffilmark{\ref{UW}},
Neta A. Bahcall\altaffilmark{\ref{Princeton}},
Robert H. Becker\altaffilmark{\ref{Davis}},
Mariangela Bernardi\altaffilmark{\ref{Chicago}},
Michael Blanton\altaffilmark{\ref{NYU}},
Daniel Eisenstein\altaffilmark{\ref{Arizona}},
Xiaohui Fan\altaffilmark{\ref{IAS}},
Douglas Finkbeiner\altaffilmark{\ref{Princeton}},
Kristian Finlator\altaffilmark{\ref{Princeton}},
Joshua Frieman\altaffilmark{\ref{Chicago}},
James E. Gunn\altaffilmark{\ref{Princeton}},
Pat Hall\altaffilmark{\ref{Princeton}},
Rita S.J. Kim\altaffilmark{\ref{Baltimore}},
Ali Kinkhabwala\altaffilmark{\ref{Princeton}},
Vijay K. Narayanan\altaffilmark{\ref{Princeton}},
Constance M. Rockosi\altaffilmark{\ref{Chicago}},
David Schlegel\altaffilmark{\ref{Princeton}},
Donald P. Schneider\altaffilmark{\ref{PennState}},
Iskra Strateva\altaffilmark{\ref{Princeton}},
Mark Subbarao\altaffilmark{\ref{Chicago}},
Aniruddha R. Thakar\altaffilmark{\ref{Baltimore}},
Wolfgang Voges\altaffilmark{\ref{MPIE}},
Richard L. White\altaffilmark{\ref{stsci}},
Brian Yanny\altaffilmark{\ref{Fermilab}},
Jonathan Brinkmann\altaffilmark{\ref{APO}},
Mamoru Doi,\altaffilmark{\ref{UT}},
Masataka Fukugita,\altaffilmark{\ref{UT2}},
Gregory S. Hennessy\altaffilmark{\ref{USNO1}},
Jeffrey A. Munn\altaffilmark{\ref{USNO2}},
Robert C. Nichol\altaffilmark{\ref{CMU}},
Donald G. York\altaffilmark{\ref{Chicago}}
}


\newcounter{address}
\setcounter{address}{1} \altaffiltext{\theaddress}{ Princeton University
Observatory, Princeton, NJ 08544 \label{Princeton}}
\addtocounter{address}{1} \altaffiltext{\theaddress}{ Fermi National Accelerator
Laboratory, P.O. Box 500, Batavia, IL 60510 \label{Fermilab}}
\addtocounter{address}{1} \altaffiltext{\theaddress}{Department of Astronomy
and Astrophysics, The Pennsylvania State University, University Park, PA 16802
\label{PennState}}
\addtocounter{address}{1} \altaffiltext{\theaddress}{Department of Physics and
Astronomy, The John Hopkins University, 3701 San Martin Drive, Baltimore, MD 21218
\label{Baltimore}}
\addtocounter{address}{1} \altaffiltext{\theaddress}{University of Washington,
Dept. of Astronomy, Box 351580, Seattle, WA 98195 \label{UW}}
\addtocounter{address}{1} \altaffiltext{\theaddress}{Physics Department, University
of California, Davis, CA 95616 \label{Davis}}
\addtocounter{address}{1} \altaffiltext{\theaddress}{University of Chicago, Astronomy
\& Astrophysics Center, 5640 S. Ellis Ave., Chicago, IL 60637 \label{Chicago}}
\addtocounter{address}{1}\altaffiltext{\theaddress} {Physics Department, The New
York University, 4 Washington Place, New York, NY 10003\label{NYU}}
\addtocounter{address}{1} \altaffiltext{\theaddress}{Steward Observatory,
933 N. Cherry Ave., Tucson, AZ 85721 \label{Arizona}}
\addtocounter{address}{1} \altaffiltext{\theaddress}{Institute for Advanced Study,
Olden Lane, Princeton, NJ 08540\label{IAS}}
\addtocounter{address}{1} \altaffiltext{\theaddress}{Max-Planck-Institute f\"ur
Extraterrestrische Physik, Karl-Schwarzschild-Str. 1, Postfach 1317, D-85741 Garching,
Germany\label{MPIE}}
\addtocounter{address}{1} \altaffiltext{\theaddress}{Space Telescope Science Institute,
3700 San Martin Dr., Baltimore, MD 21218 \label{stsci}}
\addtocounter{address}{1} \altaffiltext{\theaddress}{Apache Point Observatory,
2001 Apache Point Road, P.O. Box 59, Sunspot, NM 88349-0059 \label{APO}}
\addtocounter{address}{1}
\altaffiltext{\theaddress}{Institute of Astronomy, University of Tokyo, 2-21-1 Osawa,
Mitaka, Tokyo 181-0015, Japan
\label{UT}}
\addtocounter{address}{1}
\altaffiltext{\theaddress}{Institute for Cosmic Ray Research, University of Tokyo, 
5-1-5 Kashiwa, Kashiwa City, Chiba 277-8582, Japan
\label{UT2}}
\addtocounter{address}{1}
\altaffiltext{\theaddress}{U.S. Naval Observatory,
Washington, DC  20392-5420
\label{USNO1}}
\addtocounter{address}{1}
\altaffiltext{\theaddress}{U.S. Naval Observatory,
Flagstaff Station, P.O. Box 1149, Flagstaff, AZ 86002
\label{USNO2}}
\addtocounter{address}{1}
\altaffiltext{\theaddress}{Department of Physics, Carnegie Mellon University, 5000
Forbes Avenue, Pittsburgh, PA 15232
\label{CMU}}

\begin{abstract}

We discuss the optical and radio properties of \about30,000 FIRST (radio, 20 cm,
sensitive to 1 mJy) sources positionally associated within 1.5 arcsec with an SDSS
(optical, sensitive to \r\about22.2) source in 1230 deg$^2$ of sky. The matched sample
represents \about30\% of the 108,000 FIRST sources and 0.1\% of the $2.5\times10^7$
SDSS sources in the studied region. SDSS spectra are available for 4,300 galaxies and
1,154 quasars from the matched sample, and for a control sample of 140,000 galaxies
and 20,000 quasars in 1030 deg$^2$ of sky. This large and unbiased catalog of optical
identifications provides much firmer statistical footing for existing results and
allows several new findings.

The majority (83\%) of the FIRST sources identified with an SDSS source brighter than 
\r=21 are optically resolved; the fraction of resolved objects among the matched sources 
is a function of the radio flux, increasing from \about50\% at the bright end to \about90\% 
at the FIRST faint limit. Nearly all optically unresolved radio sources have non-stellar 
colors indicative of quasars. We estimate an upper limit of \about5\% for the fraction 
of quasars with broad-band optical colors indistinguishable from those of stars. 
The distribution of quasars in the radio flux -- optical flux plane  supports the existence 
of the ``quasar radio-dichotomy"; 8$\pm$1\% of all quasars with \i$<$18.5 are radio-loud 
and this fraction seems independent of redshift and optical luminosity. The radio-loud 
quasars have a redder median color by 0.08$\pm$0.02 mag, and show a 3 times larger fraction
of objects with extremely red colors.

FIRST galaxies represent 5\% of all SDSS galaxies with \r$<$17.5, and 1\% for \r$<$20,
and are dominated by red (\ur$>$2.22) galaxies, especially those with \r$>$17.5.
Magnitude and redshift limited samples show that radio galaxies have a different optical
luminosity distribution than non-radio galaxies selected by the same criteria; when
galaxies are further separated by their colors, this result remains valid for both blue and
red galaxies. For a given optical luminosity and redshift, the observed optical colors
of radio-galaxies are indistinguishable from those of all SDSS galaxies selected by
identical criteria. The distributions of radio-to-optical flux ratio are similar for
blue and red galaxies in redshift-limited samples; this similarity implies that the
difference in their luminosity functions, and resulting selection effects, are the dominant
cause for the preponderance of red radio galaxies in flux-limited samples. The fraction
of radio galaxies whose emission line ratios indicate an AGN (30\%) rather than a starburst
origin is 6 times larger than the corresponding fraction for all SDSS galaxies (\r$<$17.5).
We confirm that the AGN-to-starburst galaxy number ratio increases with radio flux,
and find that radio emission from AGNs is more concentrated than radio emission from
starburst galaxies.

\end{abstract}


\section{                         Introduction                }

Statistical studies of the radio emission from extra-galactic sources are entering
a new era due to the availability of large sky area high-resolution radio surveys
that are sensitive to mJy levels (e.g. FIRST, Becker, White \& Helfand 1995; NVSS,
Condon {\em et al.} 1998; for an informative overview and a comparison of modern
radio surveys see De Breuck {\em et al.} 2000). However, to fully utilize the
strength of these new radio surveys, the optical properties of the sources must be
determined. The Sloan Digital Sky Survey (SDSS, York {\em et al.} 2000) is a good
match in areal coverage and depth (\r\about22.2) to the new radio surveys. The SDSS is producing
five-color optical images and photometry for more than 10$^8$ Galactic and extra-galactic
sources, as well as spectra for about 10$^6$ galaxies and 10$^5$ quasars. The accurate
photometry and detailed morphological and spectroscopic information can be used to
efficiently separate sources into stars, quasars and galaxies, and to study in detail
the optical properties of radio sources.

In this work we discuss the properties of sources observed by the FIRST and SDSS
surveys. We use the FIRST survey because it has superior astrometric accuracy,
resolution and faint sensitivity limit compared to other contemporary large area
radio surveys (e.g. NVSS). We use the SDSS because it has the best photometric and
astrometric accuracy available for a large area optical survey, 5-band data
extending from 3000 \AA\ to 10,000 \AA, and spectra for a large number of
extragalactic sources. The largest previous samples of optical identifications
for FIRST sources are based on the APM survey. McMahon {\em et al.}
(2001) describe an identification program for 382,892 FIRST sources from 4150
deg$^2$ of the north Galactic cap that resulted in \about70,000 optical counterparts, and
Magliocchetti \& Maddox (2001) present a detailed analysis for \about4000 sources.
The sample discussed here, with SDSS identification for \about 30,000 FIRST
sources, has the advantage of more than five times more accurate optical astrometry
and photometry; it is also about 1 magnitude deeper, and utilizes 5-color, instead
of 2-color, optical information. In addition, spectra are available for a subset of
5,454 matched sources.

A brief description of the FIRST and SDSS data is provided in \S 2. We describe the
matched data and the basic matching statistics in \S 3. A more detailed study of
the optical and radio properties of quasars and galaxies is presented in \S 4 and
\S 5, and we discuss our results in \S 6.

\section{                             The Surveys                       }

\subsection{                    Sloan Digital Sky Survey                 }

\label{SDSSbasics}

\subsubsection{                 Technical Summary                        }

The SDSS\footnote{See http://www.astro.princeton.edu/PBOOK/welcome.htm} is a
digital photometric and spectroscopic survey which will cover one quarter of the
Celestial Sphere in the North Galactic cap and produce a smaller area ($\sim$ 225
deg$^2$), but much deeper, survey in the Southern Galactic hemisphere (York {\em
et al.} 2000; Stoughton {\em et al.} 2002, hereafter EDR, and references
therein). The flux densities of detected objects are measured almost simultaneously
in five bands ($u$, $g$, $r$, $i$, and $z$, \cite{F96}) with effective
wavelengths of 3551 \AA, 4686 \AA, 6166 \AA, 7480 \AA, and 8932 \AA\ (Gunn {\em et
al.} 1998). The resulting catalog is 95\% complete\footnote{These values are
determined by comparing multiple scans of the same area obtained during the
commissioning year. Typical seeing in these observations was 1.6$\pm$0.2 arcsec.}
for point sources to limiting AB magnitudes of 22.0, 22.2, 22.2, 21.3, and 20.5 in
the North Galactic cap\footnote{We refer to the measured magnitudes in this paper
as $u^*, g^*, r^*, i^*,$ and $z^*$ because the absolute calibration of the SDSS
photometric system is still uncertain at the $\la 0.05^m$ level. The SDSS filters
themselves are referred to as $u, g, r, i,$ and $z$. All magnitudes are given on
the AB$_\nu$ system (Oke \& Gunn, 1983); for additional discussion regarding the
SDSS photometric system see Fukugita {\em et al.} (1996).}, in $u$, $g$, $r$, $i$,
and $z$, respectively. The eventual survey sky
coverage of about $\pi$ steradians (10,000 deg$^2$) will result in photometric
measurements to the above detection limits for about 100 million stars. Astrometric
positions are accurate to about 0.1 arcsec (rms per coordinate) for sources brighter
than 20.5$^m$, and the morphological information from the images allows robust star-galaxy
separation down to \r $\sim$ 21.5$^m$ (Lupton {\em et al.} 2002). The spectra have
a resolution of 1800-2000 in the
wavelength range from 3800 to 9200 \AA. Extragalactic sources targeted in the SDSS
spectroscopic survey include a flux-limited ``main" galaxy sample (\r$<$17.77, Strauss
{\em et al.} 2002), the luminous red galaxy sample (Eisenstein {\em et al.} 2002,
hereafter E02), and quasars (Richards {\em et al.} 2002). Further technical details
about the SDSS data can be found in EDR and references therein.

\subsubsection{  SDSS color-color and color-magnitude diagrams      }
\label{sdssccd}

The position of an object in SDSS color and magnitude space can be used to
constrain its nature, thus providing an efficient method to analyze the properties
of optically identified radio sources. The color-color and color-magnitude diagrams
which summarize the photometric properties of SDSS sources are shown in Figure
\ref{SDSS3}. In this and all other figures, we correct the optical magnitudes for interstellar
extinction, determined from the maps of Schlegel, Finkbeiner \& Davis (1998).
Typical $r$ band absorption values ($A_r = 0.84 A_V$) for the high-latitude regions
discussed in this work are 0.05 to 0.15. Throughout this work we use ``model"
magnitudes\footnote{Note that the SDSS photometric system uses asinh magnitudes
(\cite{LGS99}).}, as computed by the photometric pipeline ({\it photo}, version
v5\_2, for details see \cite{Lupton02} and EDR). The model magnitudes are
measured by fitting an exponential and a de Vaucouleurs profile of arbitrary
inclination to the two-dimensional image and convolved with the local
point-spread function, and using the formally better model in the $r$ band to evaluate
the magnitude. Photometric errors are typically 0.03$^m$ at the bright end
($r^* < 20^m$), and increase to about 0.1$^m$ at $r^* \about 21^m$, the faint limit
relevant in this work (for more details see Ivezi\'c {\em et al.} 2000 and Strateva
{\em et al.} 2001, hereafter S01).

The top left panel in Figure \ref{SDSS3} displays the \gr\ vs. \ug\ color-color
diagram for $\about$ 300,000 objects with errors less than 0.1 mag in the plotted
bands, observed in 50 deg$^2$ of sky during the SDSS commissioning phase. The
unresolved sources are shown by dots, and the resolved sources by linearly spaced
contours. The low-redshift quasars ($z$ $\la$
2.5), selected by their blue \ug\ colors indicating UV excess (hereafter UVX,
roughly in the region $-0.6 < \ug <$ 0.6, $-0.2 < \gr <$ 0.6), are shown as circles
(for more details about SDSS quasar targeting strategy see Richards {\em et al.} 2002).
Most of the unresolved sources are stars.  The position of a star in color-color
diagrams is mainly determined by its spectral type (Finlator {\em et al.} 2000,
hereafter F00, and references therein).
For most of its length, the locus in the \gr\ vs. \ug\ diagram consists of
stars with spectral types ranging from F to late K, with late K and M stars
distributed at the red end of the stellar locus, as marked in the figure (in all
color-color diagrams red is towards the upper right). Different M spectral subtypes
cannot be distinguished in the \gr\ vs. \ug\ diagram, and are better separated in
the \ri\ vs. \gr\ diagram (the top right panel in Figure \ref{SDSS3}), where they
occupy the vertical part of the stellar locus with \gr $\about$ 1.4. The modeling
of the stellar populations observed by SDSS (F00) indicates that the vast majority
of these stars (about 99\%) are on the main sequence.

The lower two panels in Figure \ref{SDSS3} display the color-magnitude diagrams for
unresolved (left) and resolved (right) sources. Blue stars (\gr$<$1) are more luminous
 than red stars (\gr$>$1)
and are thus observable to comparatively larger distances (about 1-10 kpc for blue
stars vs. 0.1-1 kpc for red stars). This, plus the steeply increasing luminosity
function for the less luminous red stars gives rise to the observed bimodal stellar
distribution (for details see Chen {\em et al.} 2000). The distribution of the \gr\ color
for galaxies is very narrow ($\la$ 0.2 FWHM) for \r$\la$16, and widens considerably
for fainter objects; the width of the \gr\ distribution
is significantly larger than the photometric errors (1 mag vs. $<$0.05 mag), and
thus represents a real spread in colors. The asymmetric distribution
suggests that there are at least two galaxy populations.

The distribution of galaxies in SDSS color-color space has been studied by
Shimasaku {\em et al.} (2001) and S01. S01 found that galaxies show a strongly
bimodal distribution of \ur\ color (also visible in the upper left panel in Figure
\ref{SDSS3}), and demonstrated that the two groups can be associated with
late type (blue group) and early type (red group) galaxies. However, faint
red galaxies (which dominate the radio galaxy sample) have poorly determined
$u$-band magnitudes, not allowing reliable galaxy type separation on the basis of
color. Thus, here we separate the two classes in the \r\ vs. \gr\ color-magnitude diagram.
Figure \ref{GALcmd} shows such diagrams for \about190,000 galaxies from \about100
deg$^2$ of sky, separated according to the \ur\ criterion proposed by S01: the
upper panel shows \about 126,000 galaxies with \ur $<$ 2.22, and the lower panel
shows \about64,000 galaxies with \ur $>$ 2.22.
Figure \ref{GALcmd} demonstrates that an approximate separation of the two basic
galaxy types is possible even when only the \g\ and \r\ band magnitudes are used. The
mixing of the two types in the central region (IIb) of the diagram is probably a
consequence of the fact that the $u$-band magnitudes become less accurate at the
faint end, but K correction effects may also play a role.

The dashed lines in Figure \ref{GALcmd} divide the \r\ vs. \gr\ color-magnitude
diagram into several characteristic regions that are a convenient analysis tool
when studying various subsamples of galaxies. Galaxies brighter than \r = 17.5
(close to the limit of \r=17.77 for
the SDSS main spectroscopic galaxy sample, see Strauss {\em et al.} 2002) display a
very narrow \gr\ color distribution. We divide this magnitude range into a region
including the core of the distribution (Ib), and the two regions bluer (Ia) and
redder (Ic) than the core.

The regions with 17.5 $<$ \r $<$ 20 are defined by following the boundaries of the
distribution of the two galaxy types: the region IIa is dominated by blue
galaxies, while the regions IIc and IId are dominated by red galaxies. The IIb
region contains substantial fractions of both galaxy types. The distinction between
IIc and IId regions is made, somewhat arbitrarily, by extending the separation
between the Ib and Ic regions parallel to the boundary between the IIb and IIc
regions. The definition of region IId is practically identical to the
SDSS spectroscopic targeting boundary for luminous red galaxies (E02).
Table 1 shows the counts of galaxies per unit solid angle (deg$^{-2}$), and
the fraction of blue (\ur $<$ 2.22) and red (\ur $>$ 2.22) galaxies in each region.

\subsection{                           FIRST Survey                           }

The FIRST survey (Faint Images of the Radio Sky at Twenty-Centimeters, Becker,
White \& Helfand 1995) is using the Very Large Array to produce a map of the 20 cm
(1.4 GHz) sky with a beam size of 5.4 arcsec and an rms sensitivity of about 0.15
mJy/beam\footnote{See
http://sundog.stsci.edu}. The survey will cover an area of about 10,000 deg$^2$ in
the north Galactic cap and a smaller area along the Celestial equator,
corresponding to the sky regions observed by SDSS. With a source surface density
of \about90 deg$^{-2}$, the final catalog will include \about10$^6$ sources. The
survey is currently 70\% complete. At the 1 mJy source detection threshold, about
35\% of FIRST sources have resolved structure on scales from 2--30 arcsec.

The FIRST catalog lists two types of 20-cm continuum flux density: the peak value,
\Fpeak, and the integrated flux density, \Fint. These measurements are derived from
fitting a two-dimensional Gaussian to each source, where the source maps are
generated from the coadded images from twelve pointings. For convenience, we define
an ``AB radio magnitude"
\eq{
            t = -2.5 \log \left({F_{int} \over 3631 \, {\rm Jy}}\right),
}
which places the radio magnitudes on the AB$_\nu$ system of Oke \& Gunn (1983).
One of the advantages of that system is that the zero-point (3631 Jy) does not 
depend on wavelength.

To quantify the radio morphology, we define a dimensionless measure of concentration
\eq{
                  \theta =  \left({F_{int} \over F_{peak}}\right)^{1/2}.
}
Sources with resolved radio emission will have $\theta > 1$.

The FIRST sensitivity limit of 1 mJy (for the peak flux density) corresponds to
$t=16.4$; we used the total flux, $F_{int}$, to define the $t$ magnitude. The various
spectral indices (defined by $F_\nu \propto \nu^\alpha$) that can be formed between
an SDSS wavelength with a measured magnitude $m$, and the FIRST wavelength are then
\eq{
  \alpha_{tm} = { 0.4  \over \log(20 \, {\rm cm} / \lambda_m)} \, (t-m),
}
where $\lambda_m$ is the effective wavelength corresponding to $m$. For example,
for the $i$ band ($\lambda_i = 7480$ \AA)
\eq{ \label{alpha}
              \alpha_{ti} = 0.0737 \, (t - \i).
}
We will also find it useful to define the ratio of the radio to optical flux density
(without including the K correction\footnote{Throughout this paper we use the term
``K correction" as it was defined by Schneider, Gunn \& Hoessel (1983).}) as
\eq{
\label{Rmt}
            R_m = \log(F_{radio}/F_{optical}) =  0.4 \, (m - t),
}
where $m$ is one of the SDSS magnitudes. In this work we use $R_r$, $R_i$, and $R_z$.
Note that some papers define $R$ without the logarithm.

\section{               The Matched Data and the Matching Statistics      }

Here we present a brief summary of SDSS and FIRST data used in this work.
We also describe the matching algorithm, discuss the star/galaxy separation,
and analyze the radio differences between matched and unmatched sources,
and between optically unresolved and resolved matched sources.

\subsection{                       SDSS Data                           }
\label{SDSSdata}

We utilize SDSS imaging and spectroscopic observations that were reduced and
calibrated prior to October 8, 2001, and that overlap the area already scanned by
the FIRST survey. The imaging data cover a 1230 deg$^2$ large area on the sky and
include $2.53 \times 10^7$ unique unsaturated (\r$>$14) SDSS sources. The distribution
of a sparse sample of these sources on the sky is shown in the top panel in Figure
\ref{footprint}. The spectroscopic data are available for a 774 deg$^2$ (63\%)
subregion and include spectra for $1.21 \times 10^5$ objects. The sky
distribution of a sparse sample of sources with spectra is shown in the bottom
panel in Figure \ref{footprint}. To test the robustness of analyzed quantities regarding
the choice of area on the sky, we also use a smaller subsample based on four SDSS
commissioning runs (94, 125, 752 and 756) taken during the Fall of 1998 and the
Spring of 1999. These data are part of the publicly available SDSS Early Data
Release\footnote{Accessible from http://www.sdss.org} (see EDR) and include
$6.68 \times 10^6$ unique unsaturated objects in a 325 deg$^2$ large region of sky
bounded by $-1.25^\circ < \delta_{2000} < 1.25^\circ$ and either $0^h$ $40^m$ $<$
$\alpha_{2000}$ $<$ 3$^h$ 20$^m$ (runs 94 and 125), or $9^h 40^m$ $<$
$\alpha_{2000}$ $<$ 15$^h$ 40$^m$ (runs 752 and 756). Spectra are available for
38,000 objects from this smaller data set (the EDR sample hereafter).

Throughout the paper we introduce and describe various subsamples designed to avoid
selection effects. A summary of these subsamples is provided in Appendix A.

\subsection{                        FIRST Data                           }
\label{first}

The full area (1230 deg$^2$) analyzed here includes 107,654 FIRST sources. The top
panel in Figure \ref{rawFIRST} shows the $\log(\theta^2)$ vs. $t$ diagram for the
28,476 sources in the EDR sample region. The radio emission from the sources above the
$\log(\theta^2) = 0$ line is resolved. The diagonal cutoff running from the top to
the lower right corner is due to the FIRST faint limit; low surface brightness
sources (i.e. those that are large and faint) are not included in the catalog.
The differential $t$ distribution
(``counts") for all sources is shown in the bottom panel with circles, and
separately with triangles for the 9,823 sources with $\log(\theta^2) > 0.1$ (note
that this condition reliably selects resolved radio sources only for $t \la 15$
because noise affects the fainter sources). The counts suggest that the FIRST
sample is complete for $t \la 15.5$. The differential counts (mag$^{-1}$ deg$^{-2}$)
of all FIRST sources in the $11.5 < t < 15.5$ range can be described by
\eq{
                   \log(n) = -3.12 + 0.31 \, t.
}
This fit is shown by the dashed line and agrees well with the results discussed
by Windhorst {\em et al.} (1985, for related discussion see also White {\em et
al.} 1997). The slope of the log($n$) -- $t$ relation for sources with
$\log(\theta^2) > 0.1$ in the same $t$ range is statistically indistinguishable
from the slope for the whole sample.

\subsection{     The Positional Matching of SDSS and FIRST Catalogs }

We first positionally match all sources from both catalogs whose positions
agree to better than 3 arcsec, and find 37,210 such pairs\footnote{The matching of
core-lobe and double-lobe sources is discussed in Section \ref{discussion}.}. The
distribution of the distance between the SDSS and FIRST positions, $d$, for the 10,084
pairs from the EDR subsample is shown in the top panel in Figure \ref{offset}.
In order to test whether the distance distribution depends on optical morphology,
we split the EDR sample into 1,999 optically unresolved and 8,085 resolved
sources (for a discussion of star/galaxy separation see Section
\ref{SGclassification}). As evident, the two distributions\footnote{All histograms
marked as n/N$_{tot}$ are normalized such that the area under the curve is unity.}
are similar.

The increase in the number of matches with $d \ga 2.5$ arcsec is consistent with
expected random associations given the number density of FIRST and SDSS sources
(for details see Knapp {\em et al.} 2002, hereafter K02). Based on this histogram,
we choose 1.5 arcsec as the limiting distance for a match to be considered as an
optical identification, and find 29,528 matches satisfying this criterion.

This cutoff is a trade-off between the completeness and contamination of the
sample. For a cutoff at 3 arcsec practically all true matches (estimated to be
33,800 after subtracting the estimated number of random matches) are included in
the sample, but the contamination from random matches is roughly 9\%. On the other
hand, a cutoff at 1 arcsec, with a contamination of 1.5\%, is only 72\% complete.
The chosen cutoff results in a 85\% complete sample with a contamination of 3\%.
The high completeness and low contamination are due to the excellent astrometric
accuracy of both SDSS and FIRST. As a comparison, Magliochetti \& Maddox (2001)
used a 2 arcsec cutoff for the APM-FIRST matches, and Sadler {\em et al.} (2002)
used a 10 arcsec cutoff for the NVSS-2dFGRS matches.

Based on statistical considerations, the 29,528 optical identifications
include \about28,684 true associations and \about 844 random matches.  The
estimated completeness implies that for the 107,654 FIRST sources there are 33,746
SDSS counterparts, or 31\% of all FIRST sources\footnote{The fraction of optically
identified FIRST sources depends to some extent on SDSS observing conditions,
particularly on seeing which determines the SDSS imaging depth.} (of course, due to
the completeness vs. contamination trade-off, robust identifications can be made
only for 27\% of FIRST sources). These identifications represent \about0.14\% of all
SDSS sources in the analyzed region.

\subsection{     The Astrometric Accuracy of SDSS and FIRST Catalogs }

The sample discussed here is sufficiently large to determine systematic
astrometric offsets between SDSS and FIRST catalogs. The middle and
bottom panels in Figure \ref{offset} show the astrometric offsets in each
equatorial coordinate for sources brighter than \r=20 and \t=15, that
are least affected by measurement noise.
These histograms show a 0.045 arcsec offset in right ascension and
0.120 arcsec offset in declination. Systematic offsets in the SDSS
astrometric calibrations are thought not to exceed 0.020-0.030 arcsec
(Pier {\em et al.} 2002). An additional 0.020 arcsec systematic error is
present in {\em photo} v5\_2 astrometry due to the use of different centroiding
algorithms in different pipelines (this will be eliminated in the next
version of the pipelines and all subsequent data releases). Thus, at most
0.050 arcsec of the offset may be attributable to systematics in the SDSS
astrometry, implying similar systematic errors in the FIRST astrometry.
This is an excellent agreement; for comparison, the FIRST and APM
astrometric reference frames are offset by 0.8 arcsec (Magliochetti {\em
et al.} 2000).

The equivalent Gaussian widths determined from the interquartile range
($q_{75}-q_{25}=1.335\sigma$) are 0.25 arcsec for unresolved sources, and
0.35 arcsec for resolved sources (per coordinate, mean for both coordinates).
As the multiple SDSS commissioning observations of the same area show that
the positions of SDSS sources are reproducible to better than 0.10 arcsec
rms per coordinate, the implied astrometric accuracy of the FIRST catalog
is thus \about0.3 arcsec per coordinate for sources with \t$<$15. When no limit
on radio flux is imposed, the FIRST astrometric accuracy is \about0.4 arcsec
per coordinate. This is consistent with the FIRST claim that ``the individual
sources have 90\% confidence error circles of radius $<$ 1 arcsec at the survey
threshold" (Becker, White \& Helfand 1995).

\subsection{             Matched   vs.  Unmatched  FIRST  Sources        }
\label{Sec-matches}

Approximately 69\% of FIRST sources do not have an SDSS counterpart within 3
arcsec. Since the multiple SDSS scans of the same area, as well as matching to the
2MASS PSC sources (F00, Ivezi\'{c} {\em et al.} 2002) show that the SDSS
completeness is better than 90\% for \r $<$ 22 (and approaching 99.3\% for
\r$<$17.5), the majority of unmatched FIRST sources are probably too optically
faint to be detected in SDSS images. This conclusion is supported by deep imaging
of a 1.2 deg$^2$ region in Hercules by Waddington {\em et al.} (2000). They
identified 69 out of 72 FIRST sources from that region; all identified sources
have \r $\la$ 26, with the distribution maximum at \r \about 22. Although their
sample is small, it is the most comprehensive nearly complete sample of optically
identified radio sources at mJy flux density levels.

We find no significant differences in the radio properties between FIRST sources
with and without optical identifications. The top panel in Figure \ref{m-unmR}
compares the differential counts of FIRST sources from the EDR sample with an SDSS
counterpart within 3 arcsec, and those without, as a function
of radio AB magnitude, \t. The two lines show best fits to the counts in the
11.5 $< t < $ 15.5 range: for unmatched sources
\eq{
                          \log(n) = -3.11 + 0.30 \, t,
}
and for matched sources
\eq{
\label{NmM}
                          \log(n) = -3.66 + 0.31 \, t,
}
where $n$ is the number of sources per unit magnitude interval and square
degree. These slopes are measured with an accuracy of \about 0.02, and thus they
are statistically identical, i.e. the optical identification probability does not
depend on the radio flux for $t \ga 11$.

The bottom panel in Figure \ref{m-unmR} compares the distributions of
$\log(\theta^2)$ for the 11,817 FIRST sources from the EDR sample with $t < 15$
(the fainter sources suffer from low signal-to-noise ratio, see Figure \ref{rawFIRST}).
The number of sources decreases faster with $\theta$ for the optically identified
than for unidentified radio sources, implying that the optical identification
probability is somewhat lower for the radio resolved sources.

\subsection{             The Star-Galaxy Classification }

\label{SGclassification}

\subsubsection{           Morphological Classification                       }

The SDSS photometric pipeline classifies detected sources into resolved and
unresolved objects (see \ref{SDSSbasics} and EDR). In its current
implementation, the photometric pipeline uses a binary classification: an
object is either a ``star" (unresolved) or a ``galaxy" (resolved). Multiple
SDSS scans, comparison with the HST data, and the distribution of sources in
color-color diagrams show that the star/galaxy separation is reliable to better
than 90\% for sources with \r \about 21, and to better than 95\% for sources
with 20 $<$ \r $<$ 21 (Lupton {\em et al.} 2002). This can be seen qualitatively
in the bottom two panels in Figure \ref{SDSS3}, where the color distributions
of unresolved and resolved sources are markedly different even at the faint end
(colors are {\em not} used in the classification) .

We chose the \r $<$ 21 condition to define subsamples with robust star/galaxy
separation, resulting in 18,903 sources (out of 29,528), classified as 3,225 (17\%)
unresolved and 15,683 (83\%) resolved sources. For brevity, in the remainder of this work
we will call optically resolved FIRST sources galaxies, and optically unresolved FIRST
sources quasars. While there may be some optically resolved FIRST sources which are
not galaxies (e.g. Galactic supernova remnants), or optically unresolved FIRST sources
which are not quasars (e.g. stars with radio emission, see K02 and references therein),
their numbers in the sample discussed here are expected to be insignificant.

\subsubsection{                Color Classification                     }
\label{SGclassification2}

The color distributions of optically unresolved and resolved SDSS-FIRST sources
are very different. This difference is especially large in the \rz\ color.
The top panel in Figure \ref{colorSG} shows the \r\ vs. \rz\ color-magnitude diagram
for the 29,528 optically identified FIRST sources.  It is evident that the \rz\ color is a good
separator of the two morphological types, with the optimal cut depending on the \r\
magnitude: the unresolved sources are blue and the resolved sources are red. The
separation is clean even at the faintest levels in the diagram. The bottom panel
shows the \rz\ distributions for sources with 21$<$\r$<$ 21.5.

This good correlation between the morphology and color can be used to estimate
an upper limit on the fraction of sources with incorrect morphological classification.
We assume that all quasars are blue and all galaxies are red, and interpret
sources with ``incorrect" color as missclassified. Adopting a cut \rz=1.0 for
sources with 21$<$\r$<$ 21.5, we find that 20\% of selected quasars have \rz$>$1.0
and 26\% of selected galaxies have \rz$<$1.0. Adopting the same \rz\ cut for sources
with 21.5$<$\r$<$ 22, we find that the fractions of objects with ``incorrect'' color
are still smaller than 25\%. This is a robust upper limit on the inaccuracy of the
adopted star/galaxy separation at the faint end. Of course, some of the objects with
``incorrect'' color may be correctly classified (e.g. high redshift quasars
could have \rz$>$1.0, see Richards {\em et al.} 2002).

\subsection{            The Radio Properties of SDSS-FIRST Sources      }
\label{radioprop}

The radio properties of the galaxies and quasars brighter than \r=21 are shown in
Figure \ref{radioCounts}. The top panel shows the differential counts in radio
magnitude. The two lines show best fits to the counts in the 11.5 $< t < $ 15.5 range:
for quasars
\eq{
               \log(n) = -2.24 + 0.14 \, t,
}
and for galaxies
\eq{
               \log(n) = -5.33 + 0.40 \, t,
}
where $n$ is the number of sources per unit magnitude interval and square degree.
These fits imply that for \r$<$21 the fraction of quasars in the FIRST catalog is
a strong function of the radio flux, monotonically decreasing from $\ga$ 50\% for
bright radio sources to $\la 10\%$ at the FIRST sensitivity limit. As discussed
above, the cumulative quasar fraction among the SDSS-FIRST sources is 17\%.
These results are in qualitative agreement with those of Magliochetti \& Maddox
(2001), based on the FIRST-APM matching (for earlier results see
Windhorst {\em et al.} 1985 and references therein).

The number counts vs. magnitude slope of 0.31 for {\em all} identified sources
(eq.~\ref{NmM}) is simply a mean relation resulting from the mixing of two
different populations: quasars with a slope of 0.14, and galaxies with a slope of
0.40. Since optically identified and unidentified FIRST sources have the same
number counts slope, it is plausible that the fractions of quasars
and galaxies are roughly the same for the two subsamples. To further illustrate
this point, we add the counts of quasars and galaxies, multiply them by 5.6 to
account for the fraction of sources that are matched (17.9\% of FIRST sources pass
the cuts on maximum positional discrepancy, optical brightness and robust optical
classification), and compare them to the counts of all FIRST sources. The squares
in the top panel in Figure \ref{radioCounts} show the scaled counts for optically
identified sources, and the solid line shows the best-fit to the counts of all
FIRST sources, as discussed in \S \ref{first}. The similarity of the two
distributions supports the notion that the fractions of quasars and galaxies are
roughly the same for the matched and unmatched FIRST sources (these fractions are
17\% quasars and 83\% galaxies when no radio flux limit is imposed, and 26\%
quasars and 74\% galaxies for $t<$ 15). We will return to this point in Appendix
B where we discuss the limits on the number of quasars missed in optical surveys.

The bottom panel in Figure \ref{radioCounts} displays the distributions of the radio
concentration measure $\log(\theta^2)$ for galaxies and quasars,
where we count only the sources with $t < 15$. Note that galaxies tend to have larger
radio sizes (i.e. the radio emission is resolved on scale of \about 5 arcsec)
which suggests that a significant fraction of their radio emission either
originates {\em outside} their nucleus, or that double-lobe radio emission is
resolved (for a related discussion see section \ref{SBAGNradio}). Out of 6,646 matched
galaxies with $t < 15$, there are 2486 (37\%) with
$\log(\theta^2) > 0.1$, and 3374 (51\%) with $\log(\theta^2) > 0.05$. For comparison,
out of 2133 matched quasars with $t < 15$, there are 296 (14\%) with
$\log(\theta^2) > 0.1$, and 520 (24\%) with $\log(\theta^2) > 0.05$. Thus, the
fraction of quasars with resolved radio emission is significantly lower than the
fraction of galaxies with resolved radio emission.

\section{         The Optical and Radio Properties of SDSS-FIRST Quasars     }

In this Section we analyze optical colors and counts, and the distribution
of radio-to-optical flux ratio for the 3,225 optically unresolved SDSS-FIRST
sources; spectra are available for a subsample of 1,154 objects. A control
sample of 20,085 spectroscopically confirmed SDSS quasars\footnote{
For a detailed discussion of quasar classification from SDSS data see
Schneider {\em et al.} (2002).} is used where
appropriate. We estimate the fraction of radio quasars with stellar colors, argue
that the data analyzed here support the existence of the quasar radio-dichotomy,
discuss a color difference between radio-loud and radio-quiet quasars, and
demonstrate that the slopes of optical counts vs. magnitude relations for
radio-loud and radio-quiet quasars are indistinguishable for \i$<$18.

Although the SDSS-FIRST quasars are dominated by low-redshift ($z \la$ 2.5) objects,
the sample also includes some high-redshift objects. In a sample of 462 SDSS quasars at
redshifts greater than 3.6, 17 objects (3.7\%) are detected by FIRST, representing
1.5\% of the spectroscopically confirmed SDSS-FIRST quasars. The highest redshift object
is SDSSp J083643.85+005453.3 with a redshift of 5.82 (Fan {\em et al.} 2001).

\subsection{             The Optical Colors of FIRST Quasars                 }
\label{QSOcolors}

One of the most important advantages of a radio-selected sample of quasars is that
it suffers neither from dust extinction nor confusion with stars\footnote{K02 show
that some optically unresolved SDSS-FIRST sources are genuine radio stars. However,
their number is very much smaller than the number of quasars}. Thus, such samples
can be used to estimate a fraction of quasars with stellar colors that are missed by
optical surveys such as SDSS, and an upper limit for the number of quasars with such
a large extinction that they are undetectable at optical wavelengths. Such analysis
assumes that the color distribution of radio quasars is similar to the distribution
for the whole sample, and, in particular, that the fraction of radio quasars with
stellar colors is representative of the whole sample. Although we show in
Section~\ref{QSOcolors2} that the color distribution of radio quasars is different
from that for the whole sample, the difference is sufficiently small that it does not
significantly affect the conclusions of this section.

\subsubsection{       The Fraction of Quasars with Stellar Colors             }
\label{QSOcolors1}

The majority of optically unresolved SDSS-FIRST sources have non-stellar colors
(for a discussion of quasar colors in the SDSS photometric system see Richards
{\em et al.} 2001). We determine the fraction of sources with colors
indistinguishable from those of stars using the following procedure\footnote{Since the
sample discussed here has a fainter optical flux cutoff than objects targeted for
SDSS spectroscopy, we decided not to use the quasar targeting pipeline (Richards
{\em et al.} 2002) to define the stellar locus, because it is tuned for sources with 
\i $<$ 19 for redshifts below three.}. First we define a flux-limited sample of 
2,318 optically unresolved matched
objects with \r $<$ 20.5. This sample is then divided into three subsamples that
have all four, three, and only two reliable SDSS colors due to noise at the faint end.
By adopting stellar locus masks\footnote{By design, the stellar locus masks used
here include some sources that the quasar targeting pipeline recognizes as outliers
from the stellar locus; that is, the conservative approach adopted here
slightly {\em overestimates} the number of sources with stellar colors.}, shown
in Figure \ref{quasarcc} by the dashed lines, we
count all sources that cannot be distinguished from stars using available colors
(for a source to be considered inside the stellar locus, it must be inside the locus
in all two-dimensional color projections). The masks allow for up to 0.15 mag. distance
from a best-fit to the stellar locus in each color-color diagram, except for the
vertical part of the mask in the \gr\ vs. \ug\ diagram where the maximum allowed
distance is 0.3 mag.

For the 1,900 objects in this sample with \u $<$ 21, all four SDSS colors are
accurate to better than \about0.1 mag. We found that 75 of these (3.9$\pm$0.5\%)
have colors indistinguishable from stars. Of the remaining 1825 objects with
non-stellar colors, 1666 show strong UV color excess (\ug $<$ 0.7). Sources with
\u $>$ 21 can be divided into 340 objects with \g $<$ 21 and 78 objects with
\g $>$ 21. From the \ri\ vs. \gr\ and \iz\ vs. \ri\ color-color diagrams we
found that 179 of the former have stellar colors, and by using the \iz\ vs. \ri\
color-color diagram we found that 49 of the latter have colors indistinguishable
from stellar. In summary, 303 objects (13$\pm$1\% of the sample) cannot be
distinguished from stars by using colors alone; for objects with \u $<$ 21, this
fraction is 4\%. We obtain consistent results for a subsample with \i $<$ 19
(the SDSS spectroscopic targeting cutoff for low-redshift quasars), and when we
consider only the EDR subsample.

To illustrate these cuts, the upper left panel in Figure \ref{quasarcc} shows the \u\
vs. \ug\ color-magnitude diagram for stars and for the 537 optically unresolved
radio sources brighter than \r=20.5 from the EDR sample. Of these, 383 sources have
UV excess (\u $<$ 21 and \ug $<$ 0.7), 86 sources have non-stellar colors but without
UV excess, and 68 sources have colors indistinguishable from stellar.

The surface density of stars to a magnitude limit of \r $<$ 20.5 is more than
hundred times higher than the surface density of quasars, and consequently the
random associations are dominated by stars. The probability of finding an
unresolved SDSS source with \r $<$ 20.5 within 1.5 arcsec from a random position
is $9\times10^{-4}$ (for high galactic latitudes discussed here, see K02). Given
the number of FIRST sources (107,654), the expected number of random
associations is 97, implying that 206 (=303-97) objects (11\% out of 1900-97=1803
objects) are true optical-radio associations with stellar colors. These results
imply that the completeness of the SDSS quasar spectroscopic survey is at least
\about89\%. This fraction could be an underestimate if some of the associated
sources with stellar colors are indeed stars with radio emission, as seems to be
the case.

Optically unresolved objects with stellar colors are targeted by the SDSS
spectroscopic survey if they are associated with FIRST objects (Richards {\em et
al.} 2002; most of the quasars discussed here were not targeted using the final
version of that algorithm, for more details see EDR). However, since the
fraction of quasars with \i $<$ 19 that are detected by FIRST is only \about 8\%
(0.94 deg$^{-2}$ vs. 12.0 deg$^{-2}$), the addition of these objects to the target
list adds only \about1\% to the completeness of the spectroscopic sample.
Nevertheless, these spectra are extremely useful for examining the nature of
targeted sources, and thus for testing the above conclusions. We visually
inspected 155 available spectra from the sample of 303 objects discussed above,
and classified them into 93 stars, 6 galaxies (compact, as determined from the
imaging data) and 56 quasars. Some of these quasars have very unusual spectra; a
few examples of quasars that were targeted only because they are FIRST sources
are shown in Figure \ref{weirdQSO}. The low fraction of quasars (36\%) indicates
that the fraction of quasars missed by the SDSS quasar spectroscopic survey due to their
stellar colors may be as low as 5\% (except for redshift range of 2.5--3 where quasar
colors mimic A stars in the SDSS system, see Richards {\em et al.} 2001). The fraction
of spectroscopically confirmed stars (60\%) is about twice as high as the expected
random association rate implying that some of these are radio stars, in agreement 
with K02.

\subsubsection{ The Color Difference Between Radio-loud and Radio-quiet Quasars}
\label{QSOcolors2}

We now compare the colors of FIRST-detected quasars to those of quasars in general.
Richards {\em et al.} (2001) noted that FIRST-detected subsample of SDSS quasars
has a larger fraction of intrinsically reddened sources than all SDSS quasars.
Here we extend their analysis to a much larger sample. We study the \gi\ color
distribution because it maximizes the wavelength baseline, while avoiding the \u\
and \z\ bands which are less sensitive than the other three bands.
In our analysis we use the quasar redshifts produced by the SDSS spectroscopic
pipelines; tests have shown that these redshifts are correct for approximately
97\% of the objects (Schneider {\em et al.} 2002).

Richards {\em et al.} (2001) demonstrated that there is a tight correlation between
the redshift and SDSS colors of quasars; this relation is clearly seen in the top
panel of Figure \ref{quasargi}. The distribution of 6,567 optically selected and
spectroscopically confirmed quasars with \i $<$ 18.5 is shown by contours; those
that are resolved (2,095) are marked by crosses (mostly found at low redshift).
The 280 FIRST-detected quasars with $R_i>1$ (radio-loud\footnote{A detailed
discussion of the radio loudness is presented in \S \ref{optradio}.}) are shown as
filled circles, and the 161 FIRST-detected quasars with $R_i<1$ (radio-quiet) are
shown as open circles. It is evident that the quasars colors vary with redshift,
as discussed in detail by Richards {\em et al.} (2001). The thick solid line shows the
median \gi\ color of all optically selected quasars in the redshift range 1--2.
We subtract this median from the \gi\ color to obtain a differential color,
hereafter called color excess. The bottom panel shows the distribution of this color
excess for 2,265 quasars in that redshift range by solid squares (without error bars), 
and for 102 radio-loud quasars by circles.

The \gi\ color-excess distribution for radio-loud quasars appears to be different
from the distribution for the whole sample. First, the mean excess for the
radio-loud subsample is redder by 0.09$\pm$0.02 mag, and the median excess
by 0.08$\pm$0.02 mag. Second, the fraction of objects with very large color excess 
($>0.4$) is larger for the radio-loud subsample; we find that 4.3$\pm$0.4\% of 
quasars have such extreme \gi\ colors, while this fraction is 14$\pm$4\% for 
the radio-loud quasars. Equivalently, the fraction of radio sources in the subsample 
of quasars with extreme \gi\ color excess (\about20\%) is 2.5 times higher than 
the corresponding overall fraction for radio-loud quasars. The inspection of other 
color-color
diagrams shows that the objects with extreme \gi\ color-excess are shifted along
the stellar locus in the \ri\ vs. \gr\ color-color diagram. However, in the \gr\
vs. \ug\ diagram they are shifted above the stellar locus, and thus are easily
distinguishable from stars (i.e. they are not missed by the SDSS quasar targeting
pipeline). In the redder bands, they are not as extreme outliers
as in the bluer bands, as noted by Richards {\em et al.} (2001). Note that the
wavelength dependence of this effect is qualitatively consistent with a reddening
due to dust extinction (though, of course, it may have other causes).

The conclusion that radio quasars have statistically different optical colors is
in agreement with an analogous difference in the distribution of spectral slopes
determined from SDSS spectra. We compute spectral indices, $\alpha$, defined
by $F_\nu \propto \nu^\alpha$, as described in Vanden Berk {\em et al.} (2002).
Figure \ref{alphaD} compares the $\alpha$ distributions for 557 radio-loud
quasars, and for 6,868 quasars without FIRST detections that are brighter than
\i=19. As evident, the $\alpha$ distribution for radio-loud quasars is skewed towards more
negative values (redder spectra). The mean and median of the distribution for
radio-loud quasars are $-0.59$ and $-0.52$, while they are $-0.45$ and $-0.41$
for the full sample (the accuracy of these estimates is \about0.02). Furthermore,
the fraction of radio-loud quasars with $\alpha < -1$ is 18\%, while the
corresponding fraction for the full sample is 8.4\%. We find no correlation between
optical and optical-to-radio spectral indices for radio-loud quasars.

\subsubsection{    The Optical Counts of Radio-loud and Radio-quiet Quasars}
\label{QSOcolors3}

Figure \ref{quasaroptcounts} shows the differential counts for optically unresolved
and spectroscopically confirmed SDSS quasars from a 1030 deg$^2$ region. The turnover
at \i \about19 is due to the flux limit for spectroscopic targeting. The best fit to
these counts in the 15.5 $< \i < $ 18.0 range is
\eq{
\label{Nall}
                \log(n) = -15.15 + 0.87 \, \i,
}
where the counts are expressed per unit magnitude and per square degree. For
illustration, an approximate estimate of the quasar counts to \i \about20 is
obtained by photometric selection of unresolved SDSS sources brighter than \u=21
that show UV excess (for clarity, displayed only for \i$>$17.5). These counts
turn over for \i$\ga$20 due to the \u $<$ 21 selection cutoff\footnote{The quasar
counts from deep optical surveys do flatten for \i $\ga$ 20, and the slope becomes
\about0.3 for \i$\la$22 (Pei, 1995).}.

The optical counts of FIRST-detected quasars are affected by the 1 mJy radio flux
cutoff. This effect can be removed by imposing a sufficiently large requirement on
the value of the radio-to-optical {\em flux ratio}, so that the radio flux cutoff
becomes inconsequential. We adopt a condition $R_i>1$ which selects 969 radio-loud
quasars. Their optical counts are shown in Figure \ref{quasaroptcounts};
the best fit in the 15.5 $< \i < $ 18.0 range is
\eq{
\label{Nradio}
                \log(n) = -16.35 + 0.89 \, \i.
}

The number counts vs. magnitude slopes given by eqs. \ref{Nall} and \ref{Nradio}
are measured with an accuracy of \about 0.03, and thus they are statistically
identical, i.e. the fraction of radio-loud quasars is {\em not} a function of
optical magnitude. The SDSS and FIRST data show that the fraction of quasars
with \i$<$18.5 which are radio-loud is 8$\pm$1\% (the SDSS-FIRST sample includes 441
spectroscopically confirmed quasars with \i$<$18.5 in 774 deg$^2$ of sky, and 280 of
those have $R_i>1$; the control sample includes 4,472 spectroscopically confirmed
quasars with \i$<$18.5 in 1030 deg$^2$ of sky).

Without a restriction on the radio-to-optical flux ratio, the optical counts of
FIRST-detected quasars have a flatter slope because at the bright optical
magnitudes the FIRST survey also detects radio-quiet quasars. The counts of all
1,154 FIRST-detected quasars from a 1230 deg$^2$ region are shown in Figure
\ref{quasaroptcounts} as dots, and the best fit is
\eq{
                      \log(n) = -8.91 + 0.47 \, \i.
}
This result is in agreement with the optical counts of quasars discovered by the
FIRST Bright Quasar Survey (White {\em et al.} 2000). For \i$<$18.5, the cumulative
fraction of FIRST-detected quasars is 13\%.

\subsection{                    The Quasar Radio-Dichotomy                       }
\label{optradio}

There is controversy in the literature about the existence of a bimodality in the
distribution of radio loudness\footnote{Two definitions of radio loudness are found 
in the literature. Here we use the radio-to-optical flux ratio to quantify radio
loudness; the alternative approach based on radio luminosity is discussed in
Appendix C.} of quasars. Strittmatter {\em et al.} (1980) pointed
out that the radio-to-optical flux density ratio for optically selected quasars
appears bimodal. Many other studies found similar results, e.g. Kellermann
{\em et al.} (1989), Miller, Peacock \& Mead (1990), Stocke {\em et al.} (1992),
Hooper {\em et al.} (1995), Serjeant {\em et al.} (1998), and references therein.
However, some authors question the existence of this so-called ``radio-dichotomy",
e.g. Condon {\em et al.} (1981) and White {\em et al.} (2000). Most of these studies
are based on small samples which typically include only \about100 sources, except for
the FIRST Bright Quasar Survey (White {\em et al.}, 2000) with 600 objects.

\subsubsection{The Distribution of SDSS-FIRST Quasars in the Optical-Radio Flux plane  }

In this subsection we determine the unbiased distribution of the radio-to-optical flux
ratio, $R_i$, and argue that it supports the existence of radio-dichotomy. Following
White {\em et al.} (2000), we compute $R_i$ with observed (i.e. no K correction\footnote{There
is no difference between corrected and uncorrected $R_i$ as long as the optical and
radio spectral slopes are the same, as is often assumed ($\alpha_{opt} = \alpha_{radio}
=-0.5$).}) $t$ and \i\ magnitudes (eq.~\ref{Rmt}). To ensure that the optical and radio
fluxes are reliable, we constrain the sample to optically unresolved sources with \i $<$ 21
and $t<16.5$, resulting in 3,066 objects. Their distribution in the $t$ vs. \i\ diagram
is shown in the top panel in Figure \ref{quasartvsi}.

If the distribution of the radio-to-optical flux ratio, $R_i$, is {\em not} a
function of the optical luminosity or redshift, then the distribution of $R_i$
should be uncorrelated with the apparent optical and radio magnitudes. However,
the optical and radio flux limits (\i $<$ 21 and $t<16.5$) have a significant
effect on the {\em observed} $R_i$ distribution, and must be taken into account
properly\footnote{For example, consider a uniform distribution of points in the
$x-y$ plane that is sampled in a square defined by $0<x<1$ and $0<y<1$. The sampled
distribution of variable $\phi=y-x$ has a local maximum for $\phi=0$, although
the underlying distribution is uniform. The unbiased $\phi$ distribution in
the $-0.5 < \phi < 0.5$ range can be easily determined by considering only
the square defined by $y=x+0.5$, $y=x-0.5$, $y=-x+0.5$, and $y=-x+1.5$.
It is possible to account for the selection effects using the whole sample, as
described by e.g. Petrosian (2001).}.
The solid and four dashed lines in Figure \ref{quasartvsi} extending from the upper
left to the lower right corner show five characteristic values of $R_i$. The three
dot-dashed lines, perpendicular to $R_i=const.$ lines, define two strips in the
$t-\i$ plane that are not affected by the optical and radio flux limits for sources
with 0 $< R_i <$ 4. The $R_i$ distributions for 670 sources from these strips
are shown as the solid symbols in the bottom panel in Figure
\ref{quasartvsi}. The two histograms are statistically the same (for clarity, Poisson
error bars are shown only for one histogram) suggesting that the distribution of the
radio-to-optical flux ratio for quasars is independent of apparent optical and radio
magnitudes.

The $R_i$ distributions show a local maximum at $R_i \about 2.8$. Given that the majority 
of SDSS quasars (\about90\%) with \i$<$18.5 are not detected by FIRST, the unobserved part 
of the $R_i$ distribution must continue to rise for $R_i < 0$. In principle, the unobserved 
part of the $R_i$ distribution could be a monotonically decreasing tail extending far to 
negative values. However, deep radio studies of smaller samples (e.g. Kellermann {\em et al.} 
1989) indicate that nearly all optically bright radio-quiet sources have at most
a factor \about1000 weaker radio emission than radio-loud sources ($-2 \la R_i \la 0$). 
This implies a local minimum in the $R_i$ distribution which appears to be in the region 
0 $\la R_i \la $ 1. Thus, the $R_i$ distribution observed for SDSS-FIRST quasars, in particular 
its rise between $R_i\about1$ and $R_i\about3$, supports the existence of the radio-dichotomy. 
The loud/quiet division line at $R_i\about 1$ is consistent with previous work (e.g. Urry 
\& Padovani 1995), and implies that every quasar detected by FIRST and fainter than \i=18.5 
is radio-loud. An unbiased estimate of the number ratio of radio-loud to radio-quiet quasars
is not possible with the available data (our estimate that 8$\pm$1\% of quasars are radio-loud 
is valid for a sample limited by optical flux, \i$<$18.5). Such an estimate could be determined 
if, for example, the data were available in the strip bounded by lines at $\t=34-\i$ and 
$\t=35-\i$; that is, if radio observations of SDSS quasars that are deeper than the FIRST 
survey were available for a sufficiently large number of sources. 

The local minimum in the $R_i$ distribution is in conflict with the
suggestion by White {\em et al.} (2000) that the bimodal distribution of the radio
properties may be spurious. However, they did not correct the observed $R_i$
distribution for selection effects. The FBQS was limited in the optical range by
the POSS-I E magnitude cutoff $E < 17.8$ (corresponding to \i $\la$ 18), and in the
radio by the FIRST sensitivity limit ($t < 16.4$). Since the quasar counts steeply
increase with optical apparent magnitude, the sample is dominated by sources near
the flux limit, and the $R$ distribution is heavily weighted by sources with $R_i$
close to $R_{cutoff} = 0.4 (i_{cutoff}-t_{cutoff}) \about 0.6$. Thus the fraction
of radio-intermediate quasars detected by the FBQS is larger than in other
surveys. To illustrate this effect, we follow White {\em et al.}, and simply
determine the $R_i$ distribution for all objects in our sample with $\i < 18$ (359),
shown by open squares in the bottom panel in Figure \ref{quasartvsi}. The counts
rise towards small $R_i$ without strong indication for a local minimum because the
sample is biased by objects around the faint cutoff\footnote{This bias may be
responsible for the increased fraction of broad absorption line quasars with
intermediate radio-to-optical flux ratios ($0 < R_i < 1$) discussed by Menou
{\em et al.} (2001). However, their sample is too small to confidently exclude the
possibility that quasars with intermediate radio-to-optical flux ratios are
more likely to have broad absorption lines.}.

The distribution of data points in the top panel in Figure \ref{quasartvsi}
illustrates why the quasar samples detected in older bright radio and optical
surveys had very different $R_i$ distributions. For example, in a radio survey
with a faint flux limit of 0.1 Jy ($t$=11.4) nearly all sources have
$R_i > 2$ (i.e. those above the $t$=11.4 line). On the other hand, quasars
detected in a bright optical survey sensitive to \i=16.5 (i.e. those to the
left from the \i=16.5 line) separate into two types: those with $R_i \ga 2$
and those with $R_i \la 1$. The more sensitive SDSS and FIRST data
sample a much larger portion of the $t$--\i\ plane, and provide an improved
estimate of the overall $R_i$ distribution.

\subsubsection{The Radio-Loudness as a Function of Luminosity and Redshift   }

Hooper {\em et al.} (1995) argued that the fraction of radio-loud objects is a
function of both optical luminosity and redshift. The sample discussed here is
sufficiently large to test this suggestion. The top panel in Figure
\ref{Mredshiftquasar} compares the distribution of 280 radio-loud quasars with \i$<$18.5
and $R_i >$ 1 to the distribution of 4,472 optically selected quasars brighter
than \i=18.5 in the absolute magnitude ($M_i$) vs. redshift plane.
The absolute magnitude is strongly correlated with redshift due to
the faint optical cutoff and steeply rising optical counts.

The middle panel compares the absolute magnitude histogram for the whole sample
and for the radio-loud subsample. The bottom panel compares the redshift histograms.
There are no significant differences between the distributions of $M_i$ and
redshift for the radio subsample and for the whole sample. The small differences
for $M_i \about -23$ and for redshift \about0.5 are due to a clump of about 10
objects and are significant only at the 1-2$\sigma$ level.

We conclude that our sample does not support the suggestion that the fraction of
radio-loud quasars depends on optical luminosity or redshift. However, we caution
that, at least in principle, the luminosity and redshift dependence could conspire
to produce no observed effect (e.g. if the radio-loud fraction decreases with redshift
and increases with luminosity) due to the strong correlation between luminosity and
redshift in a magnitude-limited sample.

\section{        The Optical and Radio Properties of SDSS-FIRST Galaxies       }
\label{ORgalaxies}

Having analyzed properties of optically unresolved objects with radio 
detections, we turn now to galaxies. 
In this Section we analyze the galaxy distribution in the space spanned
by optical and radio fluxes, optical colors, redshift, and emission line
properties. The analyzed sample includes 15,683 optically resolved sources
brighter than \r=21 that are detected by FIRST; for a subsample of 5,454 sources
SDSS spectra are available. We also use a control imaging sample of 304,147
optically resolved sources, and a control spectroscopic sample which includes
141,920 galaxies.

First we compare the optical properties of radio galaxies to the properties of
galaxies from the control samples. We find that they have different optical
luminosity distributions, while their color distributions are indistinguishable
in subsamples selected by optical luminosity and redshift. This analysis is
fully based on directly observed magnitudes, without taking into account
K corrections. The effect of the optical K correction in the relevant redshift
range ($\la 0.4$) is to dim and redden galaxies by a few tenths of a magnitude
(Blanton {\em et al.} 2000, and references therein). On the other hand, the
radio K correction, due to steep radio spectra, {\it brightens} the radio flux; 
due to this effect, the radio-detection probability increases with redshift (for 
more details see De Breuck, 2000).

SDSS spectra are used to compute the line strengths for several characteristic
emission lines that allow classification of galaxies into starburst galaxies
and AGNs (active galactic nuclei). We find that the fraction of radio galaxies
whose emission line ratios indicate an AGN rather than a starburst origin (30\%)
is 6 times larger than the corresponding fraction for all SDSS galaxies (\r$<$17.5).
The radio emission from AGNs is more concentrated than radio emission from
starburst galaxies, and the AGN-to-starburst galaxy number ratio decreases
with radio flux.

\subsection{          The Optical Properties of Radio Galaxies              }
\label{GalOptProp}

\subsubsection{          The Morphological Properties                       }

First we examine the distribution of radio galaxies in the \r\ vs. \gr\
color-magnitude diagram (where morphological types are well separated, c.f. \S
\ref{sdssccd}), and then we discuss the visual inspection of multi-color SDSS
images. Figure \ref{gal-colmag-gr} displays the \r\ vs. \gr\ color-magnitude for
SDSS-FIRST galaxies and for all SDSS galaxies. The top panel shows 19,496 SDSS-FIRST
galaxies brighter
than \r=21.5 (for illustration, in this figure we relax the condition that \r$<$21),
and the bottom panel shows 4,300 galaxies for which SDSS spectra are available
(the spectroscopic sample is practically complete for \r $\la$ 17.7 in the regions
of sky studied here). The dashed lines outline the regions discussed in \S
\ref{SDSSdata} and Figure \ref{GALcmd}.

The color-magnitude distribution of radio galaxies is markedly different from that
of SDSS galaxies as a whole; at the bright end (\r$<$17.5) the fraction of radio
galaxies does not strongly depend on color (i.e. galaxy type, S01), while at the
faint end it is strongly dependent on color. For example, in region IId the radio
fraction is \about30 times larger than in region IIa, although both regions span
the same \r\ magnitude range. Table 1 lists the surface densities of radio
galaxies, expressed as a fraction of all galaxies, for each of the seven regions
outlined in Figure \ref{gal-colmag-gr}.

Region IId includes the luminous red galaxies (LRGs), which are the reddest
galaxies in the \r\ vs. \gr\ color-magnitude diagram, due to the effects of
K-corrections in the $g$ and $r$ bands (E02). Despite
being among the most distant SDSS galaxies (redshifts up to $\la0.55$), a
strikingly large fraction of LRGs (\about 3\% in region IId, see Table 1) have
radio counterparts. This effect is probably caused by steep radio spectra of these
(comparatively) high-redshift galaxies (De Breuk 2000).

We examine next the optical color images of radio galaxies. The 4,152 galaxies
with \r $<$ 17.5 are sufficiently bright and large for SDSS imaging to capture detailed
morphological information. Following S01, we first divide them into 1,153 blue galaxies
with \ur $<$ 2.22 and 2,999 red galaxies with \ur $> 2.22$ (for bright galaxies the
\ur\ color-based separation is more robust than \gr\ classification, S01). Visual
inspection of 1084 multi-color ($g-r-i$ composites) SDSS images from the EDR subsample
confirms that $\ga$ 80\% of the blue radio galaxies are spiral galaxies, while
$\ga$ 90\% of the red radio galaxies are elliptical galaxies, in agreement with the
results for the full sample (S01). We found only a few examples of clearly blue elliptical
galaxies.

The visual inspection indicates that the incidence of merging galaxies and galaxies
with disturbed structure among the blue radio galaxies is higher than for a random
sample of blue galaxies. A detailed quantification of this effect will be presented in
a future publication. For illustration, Figure \ref{colormosaic} shows composite
$1\times1$ arcmin $g-r-i$ images of 12 radio galaxies, selected to include interesting
examples. The radio position is marked by a cross. The merging nature can be discerned
more easily in Figure \ref{rmosaic} which shows the same galaxies with a stretch chosen
to emphasize the galactic nuclei. Obvious examples of mergers are shown in the top and
bottom middle panels.

\subsubsection{          The Optical Colors of Radio Galaxies            }

We now compare the optical colors of radio galaxies and other galaxies.
We demonstrate that optical colors of radio galaxies are similar to those
of all galaxies selected by the same redshift and optical luminosity criteria,
despite being markedly different in flux-limited samples.

The two top panels and the bottom left panel in Figure \ref{GALccd} compare the
distributions of SDSS galaxies and 4,152 radio galaxies brighter than \r=17.5
in color-color diagrams. The radio
galaxies appear to have redder color distributions than the full sample. This
is a selection effect; in a flux limited sample radio galaxies tend to have
larger redshifts than all galaxies as a whole (demonstrated further below).
Because of larger redshifts and optical K correction reddening (which we did
{\em not} take into account), their {\it observed} colors are redder (the evolutionary
effects may also contribute to this reddening). When selected from the same
redshift range, the colors of radio galaxies are similar to those of
other galaxies. This point is illustrated in Figure \ref{GALccd} (the bottom right
panel) where we compare the redshift vs. \ri\ color distributions for all galaxies
with spectra; radio galaxies follow the same distribution as all
SDSS galaxies with spectra.

The tendency that the colors of radio galaxies in a flux limited sample are biased
towards the red end affects the classification based on observed \ur\ color. The top
left panel in Figure \ref{GALur} compares the \r\ vs. \ur\ color-magnitude distributions
for galaxies with \r $<$ 17.5 (practically a complete sample). The fact that radio
galaxies have redder
\ur\ color is more clearly visible in the top right panel where we compare the
\ur\ histograms.  The shapes of
the two distributions are similar, with the radio galaxy color distribution redder
by \about 0.3 mag. The vertical dot-dashed line at \ur = 2.22 shows the classification
boundary between blue and red galaxies at low redshifts (S01). Assuming that all radio
galaxies are redder by \about0.3 mag, the application of this condition to radio
galaxies brighter than \r = 17.5 results in \about 20\% ``blue'' galaxies
misclassified as ``red'' galaxies. Since this effect is not very strong, in the rest
of this work we retain \ur = 2.22 as the separation boundary between blue and
red galaxies.

The \ur\ color distribution of radio galaxies is somewhat redder than that
of other galaxies because radio galaxies are sampled at higher redshifts. The bottom
left panel in Figure \ref{GALur} compares the redshift vs. \ur\ distributions
for galaxies with \r $<$ 17.5. They are same\footnote{This statement is based on
the comparison of \ur\ histograms for narrow (\about 0.01-0.05) redshift slices
in the range 0.05 $<$ redshift $<$ 0.30.} to within $\sim 0.1$ mag, further
demonstrating that the intrinsic optical colors of radio galaxies are not very
different from other galaxies. The remaining difference in colors of $\sim 0.1$ mag
is due to the color-luminosity relation and differences in luminosity distributions
between radio and other galaxies, which are discussed in the next section.

The redshift distributions for galaxies bluer and redder than \ur=2.22, and
with \r $<$ 17.5, are shown in the bottom right panel in Figure \ref{GALur}. The
sharp features visible in these curves are {\em not} due to Poisson noise (the
histograms are based on a sample including \about90,000 galaxies), rather they
reflect the large scale structure in the distribution of galaxies. The redshifts of
radio galaxies tend to be larger than for other galaxies from the same flux-limited
sample, and this effect is especially pronounced for red galaxies.

In summary, the broad band optical colors of radio galaxies in a flux limited
sample are redder than for all galaxies in the sample. This effect can be
explained by the color-redshift relations; when confined to the same redshift
range, the colors of radio galaxies are similar to those of other galaxies.

\subsubsection{   The Optical Luminosity Distribution of Radio Galaxies }

We now compare the optical luminosity distribution of radio galaxies to that
of all galaxies selected from the same narrow redshift range. The top left
panel in Figure \ref{optlum} shows the \r\ vs. \ur\ color-magnitude distribution
for galaxies with \r $<$ 17.5, redshift in the range 0.08--0.12, and, for radio
galaxies, $R_r > 0.4$. The condition on $R_r$ ensures that the radio flux limit is
not relevant (all galaxies with $R_r > 0.4$ and \r $<$ 17.5 are brighter than
$t=16.5$; for a discussion of the $R_r$ distribution see the next section). The 
top right panel compares the \ur\ histograms. Despite the narrow redshift range, 
the colors of red (\ur$\ga$ 2.2) radio galaxies are \about0.1 mag
redder than other galaxies from the same redshift range. This difference is larger
than plausible K-correction variance due to the finite redshift bin size. As can be
discerned from the top left panel, this effect is instead a consequence of
a magnitude-color correlation: more luminous red galaxies are redder than less
luminous red galaxies\footnote{This effect is not an artifact of SDSS data; for
a similar result see for example Figure 5 in Carlberg {\em et al.} (2001).} (the
slope is \about 0.1 mag/mag). The small difference in color distributions between
radio and other galaxies is due to their different magnitude distributions: radio
galaxies tend to be brighter. This is better seen in the two middle panels where we
compare the apparent magnitude distributions for blue (left panel) and red (right
panel) galaxies; all galaxies are shown by the thin solid line, and radio galaxies by
symbols.

Both blue and red radio galaxies show a peak in their apparent magnitude distributions.
The bottom two panels shows the corresponding distributions of absolute magnitudes
(we use the same cosmological parameters as in Appendix C, and do not K
correct). As evident, red radio galaxies show excess around $M_r \about -23$, and blue
radio galaxies around $M_r \about -22$. We find that the radio fraction of red galaxies
in the $-23.4 < M_r < -22.8$ range (1.2\%) is about twice as high as the corresponding
fraction for galaxies with $-22.4 < M_r < -21.8$ (0.64\%).

Interestingly, the luminosity difference between the two peaks (\about 1 mag) is similar
to the difference in characteristic luminosity between red and blue galaxies (Blanton
{\em et al.} 2000). This suggests that the luminosity function of radio galaxies with a
given color (or type) may be tied to the luminosity function of all galaxies with the
same color.

\subsubsection{The Blue vs. Red Galaxy Distribution in the Optical-Radio Flux plane}

The distribution of radio galaxies in  apparent magnitude vs. color space
is markedly different from the overall distribution of galaxies; in particular,
the radio fraction of red galaxies is much higher than that of blue galaxies.
This effect could be caused by intrinsic differences in radio emission properties
between the two types of galaxies, or they could simply be subtle selection effects
due to different luminosity distributions, flux limits and K corrections. Here we
discuss these possibilities in more detail and provide qualitative arguments that
explain the main trends in the data.

We demonstrated in preceding sections that the redshifts of radio galaxies tend to
be larger than those of other galaxies in a flux-limited sample. The ``redshift bias"
for radio galaxies is consistent with being due to radio K correction: the observed
radio flux of a galaxy with the same optical flux and the same rest-frame radio-to-optical
flux ratio as another galaxy at a lower redshift, is larger than the observed radio flux
of the closer galaxy (for more details see De Breuck 2000). Due to the faint radio flux
limit, radio galaxies in an optical and radio flux-limited sample are thus biased\footnote{We
cannot exclude the possibility that this bias is partially caused by evolutionary effects.}
towards larger redshifts.

The same K correction effect can, at least partially, account for the different
fractions of blue and red radio galaxies since they have different redshift distributions.
Red galaxies are sampled at larger redshifts than blue galaxies (see the bottom right panel
in Figure \ref{GALur}) because the former tend to be more luminous than the
latter\footnote{Blanton {\em et al.} (2000) find that galaxies with $M_r \la -22$ are
dominated by red galaxies (see also Madgwick {\em et al.} 2002 for consistent results
based on the 2dF Galaxy Redshift Survey).}. As a consequence of the difference in redshifts, the
{\em observed} radio-to-optical flux ratio is greater for red galaxies, and thus they
are preferentially detected by FIRST. We test this hypothesis by comparing the
radio-to-optical flux ratio distributions for blue and red galaxies in narrow redshift
bins. If the distributions are significantly different, then this hypothesis must be
rejected.

Figure \ref{GALtr} shows the $t$ vs. \r\ distributions for blue (top two panels)
and red (middle two panels) galaxies brighter than \r=17.5 and $t$=16, in two redshift
ranges, 0.03--0.07 (left column), and 0.08-0.12 (right column). The three dashed
lines in the top four panels show constant radio-to-optical flux ratios $R_r$ = 0,
1, and 2, as marked. In each redshift range red galaxies are  brighter in both
the optical and the radio than blue galaxies; as a result their distributions of
radio-to-optical flux ratio are very similar. This is better seen in the two bottom
panels that show the $R_r$ distributions for objects from the strips defined by the
two dot-dashed lines in each of the four top panels (note that due to optical and
radio flux limits, the $R_r$ distributions are unbiased only in the 0 $< R_r \la 2$
range). This similarity indicates that the large differences in the observed radio
fractions of blue and red galaxies may simply be due to different luminosity functions.

Windhorst {\em et al.} (1985, and references therein) show that radio
sources with $t < 13.9$ (flux density at 1.4 GHz $>$ 10 mJy) consist predominantly
of quasars and red giant elliptical galaxies, while only a few blue radio galaxies
are seen. On the other hand, for 13.9 $< t <$ 16.5 blue radio galaxies become
increasingly important. The data displayed in Figure \ref{GALtr} confirm these
results. Furthermore, we show here that the distributions of radio-to-optical flux
ratio for blue and red galaxies are very similar in the sampled range $R_r > 0$.

\subsubsection{The Luminous Red Galaxy Distribution in the Optical-Radio Flux plane}

For further investigation of the redshift dependence of the radio-to-optical flux
ratio, we select additional subsamples from two larger redshift ranges: 0.28--0.32
and 0.38--0.42. Practically all the galaxies in these samples are luminous red
galaxies; we caution that these samples are not flux limited, but are
limited by optical luminosity and color (E02). In this
analysis we use $R_z$ instead of $R_r$ because of the large K
correction in the $r$ band. The median \iz\ color of galaxies at these redshifts is
only \about 0.1 mag redder than the median \iz\ color of nearby galaxies, indicating
that the effects of optical K correction are minimized in the $z$ band.

The distribution of galaxies from these two redshift ranges in the $t$ vs. \z\ plane,
and their $R_z$ distributions are shown in Figure \ref{GALtrX}. The bottom panel
demonstrates that $R_z$ is redshift dependent: the median $R_z$ is larger by \about0.5
for the larger redshift bin. This effect is consistent with the behavior of radio
K correction. Further, both distributions show local maxima, i.e. we detect a radio
dichotomy similar to that detected for quasars. The interpretation of this result
requires detailed knowledge of radio spectral indices and thus will not be attempted
here.

\subsection{          The Spectral Properties of Radio Galaxies                 }
\label{GalSpectra}

For practically all galaxies brighter than \r \about 17.5 (regions I in Figure
\ref{gal-colmag-gr}) the SDSS spectra are available (Strauss {\em et al.} 2002). In
addition, many of the radio galaxies fainter than that limit also have spectra
(region IId) because they are targeted for SDSS spectroscopy as the luminous red galaxies
(E02). At the time of this study, spectra were available for 4,300 SDSS-FIRST
galaxies brighter than \r=21, from 63\% of the 1230 deg$^2$ area discussed here.
The distribution of these galaxies in
the \r\ vs. \gr\ color-magnitude diagram is shown in the bottom panel in Figure
\ref{gal-colmag-gr}.

This is the largest homogeneous set of radio-galaxy spectra ever obtained. The largest
previously available sample of 757 objects was produced by Sadler {\em et al.}
(2002) who matched the 1.4 GHz NRAO VLA Sky Survey (NVSS) and the 2dF Galaxy redshift
Survey (Colless {\em et al.} 2001).
A slightly smaller sample of 557 objects was obtained by Magliocchetti {\em et
al.} (2001) who matched the FIRST catalog with the 2dF Galaxy redshift Survey.
The SDSS spectra tend to have better quality than usual spectra acquired for
redshift determination, and hence can be used to measure the spectral properties
of the galaxies.

\subsubsection{         The Visual Inspection of Spectra                 }

Before we attempted automated spectral classification described in the section,
we visually inspected spectra for matched objects from the EDR sample including 308
galaxies with \ur $<$ 2.22 and $\r < 17.5$, 112 galaxies with \ur $> 2.22$ and $\r
< 17.5$, and 320 galaxies with $\gr > 1.6$ and $\r < 19$. The last subsample
was designed to include the luminous red galaxies with redshifts ranging from
\about0.35 to \about 0.45. Examples of spectra are shown in Figures
\ref{spectra1} and \ref{spectra2}.

A substantial fraction (\about20\%) of the 308 bright blue galaxies show spectra
characteristic of starburst galaxies, and \about10\% show AGN-type spectra. The
remaining blue galaxies show spectra typical for spiral galaxies (e.g.
Kennicutt 1992). The majority of bright red galaxies show spectra characteristic
of ordinary elliptical galaxies, with about one quarter classifiable as LINERs,
and a few percent indicating the presence of AGN. The subsample of luminous red
galaxies\footnote{The luminous red galaxies are those that have large bright
absolute magnitude, as opposed to the bright red galaxies that have bright apparent
magnitude.} with FIRST detections do not have significantly different spectra from the
whole sample. There are clear cases of AGNs, though at a smaller rate than for the
bright red galaxies. However, note that spectra from this subsample have lower
signal-to-noise ratios due to a fainter magnitude limit, and thus AGN may be more
difficult to recognize. These results are in qualitative agreement with those of
Sadler {\em et al.} (2002).

\subsubsection{       The Fractions of AGNs and Starburst Galaxies           }
\label{SBAGN}

A detailed quantitative study of spectral properties of the radio-galaxies from
the SDSS-FIRST sample must be automated due to the large sample size. Here we present
a preliminary determination of the fractions of AGNs and starburst galaxies
using diagnostic diagrams based on the strengths of several emission lines. Details
of the line strength determination will be presented elsewhere (Tremonti {\em et
al.} 2002).

Galaxies with emission lines due to strong star formation can be separated from
galaxies whose emission lines originate from an AGN by using optical line diagnostic
diagrams. Such a separation is possible because AGNs have a much harder ionizing
spectrum than stars. This method was first proposed by Baldwin, Phillips \&
Terlevich (1981), and was further developed semi-empirically by Osterbrock \& de
Robertis (1985) and Veilleux \& Osterbrock (1987), and theoretically by Kewley
{\em et al.} (2001). The line diagnostic diagrams are constructed with the line
strength ratios $[OIII]/H_\beta$, $[NII]/H_\alpha$, $[SII]/H_\alpha$, and
$[OI]/H_\alpha$. We computed these ratios for galaxies with \r $<$ 17.5; the
samples with unsaturated and better than 3$\sigma$ detections for $H_\alpha$,
$H_\beta$, $[OIII(5007)]$, $[NII(6584)]$, and $[SII(6717+6731)]$ include 650 SDSS-FIRST
galaxies and 16,325 SDSS galaxies (we do not use the $[OI]$ line because its 3$\sigma$ 
requirement would significantly reduce the sample size). These subsamples represent 
26\% and 18\%, respectively, of all galaxies for which the line strength measurement 
was attempted.

The distribution of the 16,325 SDSS galaxies in the $[NII]/H_\alpha$ vs.
$[OIII]/H_\beta$ and $[SII]/H_\alpha$ vs. $[OIII]/H_\beta$ diagrams is shown in
Figure \ref{AGNs}. The distribution of galaxies shows a remarkable
structure in these diagrams, rather than a random scatter\footnote{We do not find
significant changes
in the morphology of data distribution in these diagrams for various subsamples
selected from narrow redshift and apparent magnitude bins.}; the overall
distribution is in agreement with
previous work that was based on much smaller samples (e.g. Veilleux \& Osterbrock
1987, and references therein). The dashed lines, obtained theoretically by Kewley
{\em et al.} (2001), separate AGNs from starburst galaxies. Motivated
by the distribution of galaxies in the $[NII]/H_\alpha$ vs. $[OIII]/H_\beta$ diagram,
we place the additional constraint that $[NII]/H_\alpha > -0.5$ for a source to
be classified as an AGN, to exclude a small number of low-metallicity starbursts. Requiring
the same classification in {\em both} diagrams removes only \about7\% of galaxies.
We find that 5\% of all galaxies are classified as AGNs, and 88\% as starbursts,
implying a starburst-to-AGB ratio of \about18 in the full sample.
The data for the 650 radio-galaxies are shown in Figure \ref{AGNs} as circles;
the corresponding starburst-to-AGN number ratio for radio-galaxies is 2.4, significantly
smaller than for the full sample. That is, the fraction of radio galaxies whose emission
line ratios indicate an AGN rather than a starburst origin is \about6 times larger than the
corresponding fraction for all SDSS galaxies.

\subsubsection{  The Optical Colors of Galaxies with Strong Emission Lines    }

The galaxies with strong emission lines that are analyzed here are predominantly blue.
The top panel in Figure \ref{AGNrgr} shows the \r\ vs. \ur\ distribution of the 16,325
SDSS galaxies as contours (compare to the top left panel in Figure \ref{GALur}).
Their \ur\ color distribution is shown in the middle panel, for AGNs (
long-dashed line)  and starburst galaxies (short-dashed line) separately. Note that
each curve is separately normalized. AGNs have redder \ur\ colors by \about0.6 mag than
do starburst galaxies. The bottom panel compares their redshift distributions using
analogous notation. The \ur\ color and redshift distributions for radio-galaxies
are shown by dots for starburst galaxies and by triangles for AGNs. There is no
significant difference in color and redshift distributions between radio galaxies
and the full sample. In particular, radio galaxies show similar separation of
\ur\ color between starburst and AGN types.

\subsubsection{  The Radio Properties of Galaxies with Strong Emission Lines    }
\label{SBAGNradio}
We examine next the radio properties of the two classes of radio galaxies.
The top panel in Figure \ref{radioCountsAGNvsSF} compares the differential counts
as a function of the radio magnitude, and the bottom panel displays the distributions
of the radio concentration parameter $\theta$ for sources with $t < 15$. Radio galaxies
classified as starbursts are marked by circles and AGNs by triangles. Their cumulative
counts for $t < 15$ are similar, while there are three times more starburst
galaxies than AGNs in the $15 < t < 16$ range. This increase of the starburst galaxy
to AGN number ratio as the radio flux decreases is consistent with the known differences
in their radio luminosity functions (Machalski \& Godlowski 2000, Sadler {\em et al.} 2002).
Machalski \& Godlowski detected a turnover in AGN counts at the faint radio end, 
an effect not seen by Sadler {\em et al.}. The data discussed here seem to support 
the claim by Machalski \& Godlowski.

The bottom panel in Figure \ref{radioCountsAGNvsSF} compares the distributions of the
radio concentration parameter. As evident, starburst galaxies tend to have larger
radio concentration parameter than AGNs, in agreement with the expected nuclear 
origin of AGN emission. This difference strongly supports the robustness of the AGN/starburst 
galaxy separation, which is fully based on optical spectral properties; its detection 
is possible due to our large spectroscopic sample and the good spatial resolution of 
the FIRST survey.

\subsubsection{ A Strategy for Selecting $z>0.5$ Galaxy Candidates Using SDSS and FIRST}
\label{Shighz}

The furthest galaxies targeted for SDSS spectroscopy are the luminous red galaxies
(E02). The targeting strategy is based on \gr\ and \ri\ colors as a function of magnitude
and extends down to \r\about19.5, yielding about 12 galaxies per square degree,
at redshifts up to \about0.55. Galaxies at such large redshifts are
important for studies of large-scale structure and galaxy evolution, and
their usefulness increases with redshift. The selection of galaxies at even larger
redshifts can rely only on \ri\ and \iz\ colors because their $g$ band flux is too
small to be well-measured. However, the efficiency of such color selection is low
due to large photometric errors and errors in star/galaxy separation since most of
high-redshift galaxy candidates are faint (\r $\la 21.5$). We show here that the
selection efficiency can be increased by combining SDSS and FIRST data. Since
the number of radio stars is insignificant, they are not a serious source of
contamination.

The \ri\ colors of SDSS galaxies at redshifts above \about0.4 correlate well with
their redshifts. By fitting the \ri\ color-redshift distribution shown in the
bottom right panel\footnote{Most of the galaxies at redshifts larger than \about0.55
with spectra were selected as high-redshift quasar candidates and FIRST quasar candidates (for
details see Richards {\em et al.} 2002).} in Figure \ref{GALccd}, we find that a
best-fit relation
\eq{
              {\rm  redshift} = 0.54 \, (\r-\i) + 0.02
}
produces redshifts within 0.05 (root-mean-square scatter) from the spectroscopically
measured values for 0.7 $<$ \ri $<$ 1.5 (0.4 $<$ redshift $<$ 0.8). Thus, galaxies
with redshifts in the range 0.5--0.8 can be selected by requiring 0.9 $<$ \ri $<$ 1.5.
The median \iz\ color of both spectroscopically confirmed galaxies and of SDSS-FIRST
galaxies with such \ri\ colors is \about0.5-0.6. This region overlaps with the stellar
locus in the \iz\ vs. \ri\ color-color diagram.

To illustrate the effect of increased photometric errors close to the faint end,
we show in the top two panels in Figure \ref{highz} the \iz\ vs. \ri\ color-color
diagrams for sources selected in two magnitude bins.  The top left
panel shows 20,000 sources with \r$<$17.5 from \about100 deg$^{-2}$ of sky, and the top
right panel shows \about 20,000 sources with 21.4 $<$ \r $<$ 21.5 from the same
region. The faint unresolved and resolved sources fully overlap in the
\iz\ vs. \ri\ color-color diagram.

The surface density of SDSS sources with 0.9 $<$ \ri $<$ 1.5 and 19 $<$ \r $<$ 21.5
is \about334 deg$^{-2}$ for unresolved sources, and \about65 deg$^{-2}$ for resolved
sources\footnote{Galaxies outnumber stars in SDSS data at \r\about21. However, the
opposite is true in the relevant narrow range of \ri\ color.}. The repeatability of
the star-galaxy separation is at the level of 90\% for \r \about21 and further
deteriorates at the fainter levels.  If all sources with colors appropriate for 
galaxies at $z\ga 0.5$ were targeted, the expected efficiency would be well below 50\%.

A higher targeting efficiency can be achieved for radio galaxies. As shown by
K02, the fraction of red stars detected both by SDSS and FIRST is very small,
and the condition that a candidate is detected by FIRST effectively rejects all the
stellar contaminants. The distribution of SDSS-FIRST sources from 1230 deg$^2$ of sky
in the \r\ vs. \ri\ and \iz\ vs. \ri\ diagrams is shown in the bottom two panels in
Figure \ref{highz}. The selection condition 0.9 $<$ \ri $<$ 1.5 and 19 $<$ \r $<$ 21.5 yields 3.2
sources per square degree. Thus, matched SDSS and FIRST catalogs could be
used to select a well-defined sample of 32,000 radio galaxies at redshifts in the
range 0.5--0.8 (over the eventual survey area of 10,000 deg$^2$).

Such a sample would be of great importance for studies of galaxy evolution, and for
detecting clusters of galaxies. Clusters discovered at such large redshifts provide
strong upper limits on the mass density parameter of the Universe, $\Omega_M$, and
on the amplitude of mass fluctuations, $\sigma_8$ (Bahcall \& Fan 1998, and references
therein). Such a sample could be cross-correlated with distant cluster candidates selected
by other methods, e.g. by the matched filter method (Kim {\em et al.} 2002), or surface
brightness fluctuations method (Dalcanton 1996, Bartelmann \& White 2001), thus
increasing the reliability of the matched candidates.

\section{                           Discussion                              }
\label{discussion}

This preliminary analysis of the objects detected by both SDSS and FIRST indicates
the enormous potential of combining large scale surveys at different wavelengths.
The final photometric SDSS-FIRST catalog, including five-color accurate optical
photometry and morphological information, and radio data complete to 1 mJy
level, will be available for \about 200,000 radio galaxies, and \about 40,000
radio quasars. SDSS spectra will be available for about 50,000 radio galaxies and 15,000
radio quasars; both surveys will provide outstanding astrometry ($\about 0.1$ arcsec for
SDSS and $\about 0.4$ arcsec for FIRST) for an unprecedented number of objects.
Such a large, detailed and accurate data set will certainly place studies of the
properties of extragalactic radio sources on a new level.

The main results presented here are:

\begin{itemize}

\item
We discuss optical and radio properties of \about30,000 FIRST sources positionally
matched within 1.5 arcsec to an SDSS source in 1230 deg$^2$ of sky. The matched sample
represents \about30\% of the \about108,000 FIRST sources and 0.1\% of the $2.5\times10^7$
SDSS sources in the studied region. SDSS spectra are available for 4,300 galaxies
and 1154 quasars from the matched sample.

\item
Differential radio counts of FIRST sources with and without SDSS counterparts have
indistinguishable slopes; about 25\% of FIRST sources are associated with an
SDSS source brighter than \r=21 (Section \ref{Sec-matches} and Figure
\ref{m-unmR}). This similarity, given the different number count slopes of galaxies
and quasars, suggests that the quasar-to-galaxy number ratio (\about 1:5) may be
comparable for SDSS-FIRST and FIRST-only radio sources.

\item
The majority of SDSS-FIRST sources brighter than \r=21 are optically resolved.
The fraction of resolved objects among the matched sources is a function of the
radio flux, increasing from \about50\% at the bright end to \about90\% at the
FIRST faint limit (Section \ref{radioprop} and Figure \ref{radioCounts}).

\item
Most optically unresolved radio sources have non-stellar colors indicative of
quasars. We estimate an upper limit of \about5\% for the fraction of quasars with
colors indistinguishable from those of stars, and thus missed by SDSS
spectroscopic quasar survey (Section \ref{QSOcolors1} and Figure
\ref{quasarcc}). However, a subset of those detected by FIRST are targeted for
SDSS spectroscopic observations, and will yield a large number ($\ga$ 1000) of
quasars with unusual spectra (Figure \ref{weirdQSO}).

\item
We find statistically significant differences in the optical color distribution
between radio-loud and radio-quiet quasars selected from the same redshift range:
the radio-loud quasars have a redder median color by \about0.1 mag, and show 3
times larger fraction of objects with extremely red colors (Section
\ref{QSOcolors2} and Figure \ref{quasargi}). The distributions of optical spectral
indices also indicate that spectra of radio-loud quasars tend to be redder than
spectra of radio-quiet quasars (Figure \ref{alphaD}).

\item
The fraction of optically identified quasars which are detected by FIRST decreases
with optical brightness from \about50\% for \r \about17 to \about10\% for \r
\about20; this decrease is a selection effect caused by the radio sensitivity
limit (Section \ref{QSOcolors3} and Figure \ref{quasaroptcounts}).

\item
The distribution of quasars in the radio flux -- optical flux plane
supports the reality of the ``quasar radio-dichotomy"; 8$\pm$1\% of all
quasars with \i$<$18.5 are radio-loud ($R_i > 1$), and this fraction seems independent
of redshift and optical luminosity (Section \ref{optradio} and Figures
\ref{quasartvsi} and \ref{Mredshiftquasar}).

\item
FIRST galaxies represent 5\% of all SDSS galaxies with \r$<$17.5, and 1\% for
\r$<$20, and are dominated by red galaxies, especially those with \r$>$17.5
(Section \ref{GalOptProp}, Table 1, and Figure \ref{gal-colmag-gr}). This
difference between blue and red galaxies appears to be a selection effect due to
their different luminosity functions. In particular, the distribution of the
radio-to-optical flux ratio for galaxies selected from narrow redshift bins is
indistinguishable for blue and red galaxies (Figure \ref{GALtr}).

\item
Radio galaxies have a different optical luminosity distribution than other galaxies
selected by the same redshift and optical brightness criteria; when galaxies are further
separated by their colors, this result remains valid for each color type.

\item
Radio-galaxies in luminosity and redshift-limited samples have indistinguishable
colors from other galaxies selected by identical criteria. In optical and radio
flux-limited samples radio-galaxies are biased towards larger redshifts, and thus
have redder {\em observed} colors due to optical K corrections (Section
\ref{GalOptProp}, and Figures \ref{GALccd} and \ref{GALur}).

\item

The fraction of radio galaxies whose emission line ratios indicate an AGN rather
than a starburst origin (30\%) is 6 times larger than the corresponding
fraction for all SDSS galaxies (Section \ref{GalSpectra}, and Figures \ref{AGNs}
and \ref{AGNrgr}). The AGN and starburst galaxies, classified using optical spectra,
have distinct radio properties. The AGN-to-starburst count ratio increases with
radio flux, and AGNs tend to have more concentrated radio emission than starburst
galaxies (Figure \ref{radioCountsAGNvsSF}).

\item
FIRST and SDSS data can be used to efficiently select galaxies at redshifts between
0.5 and 0.8 with a surface density of \about3 deg$^{-2}$ for candidates with
\r$<$21.5 (Section \ref{Shighz} and Figure \ref{highz}). Such a sample
would be of great importance for studies of galaxy evolution, and for detecting
clusters of galaxies.

\end{itemize}

\subsection{                   Future Work                               }

\subsubsection{       Sources with complex radio morphologies            }

In this paper we have only discussed sources for which the radio and optical positions
agree to better than 1.5 arcsec. This sample does not address the so-called core-lobe and
double-lobe radio sources (for a discussion of such sources in FIRST data see
Magliocchetti {\em et al.} 1998, and McMahon {\em et al.} 2001). We are currently
analyzing samples obtained by two additional matching methods:

\begin{itemize}
\item
If a radio source consists of two lobes separated by more than 3 arcsec,
it is not included in the sample analyzed here (double-lobe sources). We find such objects
by searching the FIRST catalog for the nearest neighbor to a FIRST source
without an SDSS counterpart within 1.5 arcsec, computing the mid-point, and
rematching these mid-points to the SDSS catalog. The initial analysis indicates
a significant excess of matches compared to the random association rate.

\item
An SDSS-FIRST source discussed here may have radio lobes that are not associated
with optical emission brighter than the SDSS faint limit (core-lobe sources). Such 
lobes can be found
by searching for nearest unmatched radio neighbors to matched SDSS-FIRST sources.
We find a significant excess of such matches compared to the random association
rate.
\end{itemize}

Detailed analysis of these two samples will be presented in a future
publication.

\subsubsection{        FIRST-only and SDSS-only sources                  }

The sample discussed here provides a good measure of the $R_i$ distribution for
radio-loud quasars. To obtain a commensurate description of the $R_i$ distribution
for radio-quiet quasars, radio observations significantly more sensitive than 
the FIRST survey are needed. Based on a sample of \about100 sources observed by 
Kellermann {\em et al.} (1989), the required sensitivity gain is about a factor 
of 10--100.

About 2/3 of the FIRST sources are not detected by SDSS. The counts of quasars
in the FIRST-only sample can be compared with deep optical counts of quasars
(\i $\la$ 25) to constrain the number of anomalously optically faint objects
missed in optical surveys (see Appnedix B). Thus, deep optical observations of regions 
containing FIRST sources without SDSS counterparts, preferably in two or more bands 
(Section \ref{SGclassification2}), would be very valuable.

\subsubsection{  Multi-wavelength observations of SDSS-FIRST sources   }

SDSS and FIRST data span a wide wavelength range, but significant parts of the
spectrum remain unexplored. The available multi-wavelength catalogs of extragalactic
sources that include X-ray,
IR, and sub-mm data are much smaller than the sample discussed here. The matching of
SDSS-FIRST sources to sensitive  large scale sky surveys at other wavelengths is
thus of obvious importance. As an example, we consider the upcoming SIRTF
observations. The SIRTF First Look Survey will observe a \about4 deg$^2$ large
region\footnote{For details see
http://sirtf.caltech.edu/SSC/T\_FLS/SSC\_FLS\_ExtrGal.html} that overlaps with the
SDSS Early Data Release (runs 1336, 1339, 1356, and 1359, see EDR). The depth
of the SIRTF First Look Survey will range from m$_{AB}$\about20 at \about4 \mic,
to m$_{AB}$\about15 at 70 \mic\, and to m$_{AB}$\about13 at 160 \mic. There are
\about150 SDSS-FIRST sources in that region, including \about20
radio-loud quasars. In addition, the SIRTF SWIRE survey will include about 35
deg$^2$ of sky that is going to be observed by SDSS (\about75\% of the overlapping
area is already observed by the SDSS imaging survey). The sensitivity of the SWIRE
survey will be 2--5 times better than that of the SIRTF First Look Survey. The
final SDSS-SIRTF-FIRST overlapping region will include \about1,500 SDSS-FIRST
sources, with \about200 radio-loud quasars; the majority of these sources are
expected to be detected by SIRTF. Such a large sample with detailed optical,
infrared and radio data will be of unprecedented size and quality.

\vskip 0.4in \leftline{Acknowledgments}

We thank Tim McKay, Paul Wiita and Moshe Elitzur for their careful reading of the
manuscript.

\v{Z}I, RHL, DS and GRK acknowledge generous support by Princeton University. MAS
acknowledges the support of NSF grant AST-0071091.
A partial support for this work was provided by NASA (to KM) through Chandra
Fellowship grant PF9-10006 awarded by the Smithsonian Astrophysical Observatory
for NASA under contract NAS8-39073.

The FIRST Survey is supported in part under the auspices of the Department of
Energy by Lawrence Livermore National Laboratory under contract No. W-7405-ENG-48
and the Institute for Geophysics and Planetary Physics.

The Sloan Digital Sky Survey (SDSS) is a joint project of The University of
Chicago, Fermilab, the Institute for Advanced Study, the Japan Participation
Group, The Johns Hopkins University, the Max-Planck-Institute for Astronomy (MPIA),
the Max-Planck-Institute for Astrophysics (MPA), New Mexico State University,
Princeton University, the United States Naval Observatory, and the University of
Washington. Apache Point Observatory, site of the SDSS, is operated by the
Astrophysical Research Consortium. Funding for the project has been provided by
the Alfred P. Sloan Foundation, the SDSS member institutions, the National
Aeronautics and Space Administration, the National Science Foundation, the U.S.
Department of Energy, the Japanese Monbukagakusho, and the Max Planck Society.
The SDSS Web site is http://www.sdss.org/.


\newpage
\clearpage

\begin{scriptsize}
\begin{deluxetable}{rrrrrr}
\tablenum{1} \tablecolumns{4} \tablewidth{360pt}
\tablecaption{
Galaxy Distribution in the SDSS \r\ vs. \gr\ Color-Magnitude Diagram$^a$. }
\tablehead {
Region & Definition$^b$ & Counts (deg$^{-2}$) & \% Blue$^c$ &  \% Red$^d$ & \% FIRST$^e$
}
\startdata
    Ia  &   \r$<$17.5 \&       \gr$<$0.7 &  24.6$\pm$0.5  &  95.0  &  5.0   &  3.9  \\
    Ib  &   \r$<$17.5 \& 0.7$<$\gr$<$1.1 &  58.0$\pm$0.7  &  16.0  & 84.0   &  4.4  \\
    Ic  &   \r$<$17.5 \&       \gr$>$1.1 &  13.2$\pm$0.3  &  2.2   & 97.8   &  8.8  \\
 all I  &   \r$<$17.5                    &  95.8$\pm$0.9  &  34.5  & 65.5   &  4.9  \\
   IIa  &    17.5$<$\r$<$20 \& \gr$<$0.7 &  317$\pm$1.8   &  97.4  &  2.6   &  0.1  \\
   IIb  &          \gr$>$0.7 \& L1$>$0   &  517$\pm$2.2   &  54.3  & 45.7   &  0.2  \\
   IIc  &          L1$<$0 \& L2$>$0      &  304$\pm$1.8   &   7.4  & 92.6   &  0.8  \\
   IId  &          L2$<$0                &  84.0$\pm$0.9  &   2.4  & 97.6   &  3.3  \\
 all II &   17.5$<$ \r$<$20              & 1222$\pm$3.5   &  50.2  & 49.8   &  0.5  \\
 I + II &     \r$<$20                    & 1318$\pm$3.7   &  49.1  & 50.9   &  0.9  \\
\enddata
\tablenotetext{a}{See Figure \ref{gal-colmag-gr}.}
\tablenotetext{b}{L1 = \r - (15.30 + 3.13*(\gr)); L2 = \r - (14.06 + 3.13*(\gr)).}
\tablenotetext{c}{Galaxies with \ur$<$2.22.}
\tablenotetext{d}{Galaxies with \ur$>$2.22.}
\tablenotetext{e}{Fraction detected by FIRST.}

\end{deluxetable}
\end{scriptsize}

\appendix{{\bf Appendix A: Summary of Data Samples Used in Analysis}}

\vskip 0.3in
\leftline{ Matched Samples }

We matched $2.53\times10^7$ SDSS sources and 107,654 FIRST sources from 1230 deg$^2$
of sky. There are 37,210 matches within 3 arcsec, and 29,528 matches within 1.5 arcsec.
Sources matched within 1.5 arcsec and brighter than \r=21 are separated into 15,683
optically resolved sources (galaxies) and 3,225 optically unresolved sources (quasars).
A subsample from SDSS Early Data Release includes 10,084 matches within 3 arcsec,
those brighter than \r=21 are separated into 1,999 quasars and 8085 galaxies.

SDSS spectra are available for a subset of 5,454 matched sources from a 774 deg$^2$
large region. The spectroscopic matched sample includes 4,300 galaxies and 1,154 quasars.

\vskip 0.2in
\leftline{ Control Samples }

The control imaging sample includes 190,577 unresolved sources (stars and quasars)
and 304,147 resolved sources (galaxies) selected from 103 deg$^2$ of sky.
The control spectroscopic sample includes 141,920 galaxies and 20,085 quasars from
a 1030 deg$^2$ large region (all available spectra at the time of writing).

\vskip 0.2in
\appendix{{\bf Appendix B: A Speculation on the Fraction of Heavily Obscured Quasars}}

\vskip 0.1in

The quasars detected by FIRST but not by SDSS may belong to the population of
anomalously optically faint quasars. The most popular argument for the existence of
such objects is the expected dust obscuration when the line of sight passes through
the optically thick torus surrounding the central engine, as advocated by the
``unified" models (e.g. Antonucci 1993 and references therein). While there are
examples of objects consistent with such an explanation (e.g. Gregg {\em et al.}
2001), it is not known how large this population is. It has been suggested that
these objects may be as populous as the optically selected quasars (Fall \& Pei 1993;
Francis, Whiting \& Webster 2000).

The data presented here suggest that the distribution of the radio-to-optical flux
ratio, $R_i$, is independent of redshift and absolute luminosity. Assuming that
the observed $R_i$ distribution is applicable to all quasars, it can be utilized,
together with optical counts of quasars from other surveys deeper than SDSS, to predict
the number of quasars detected by FIRST and not detected by SDSS. This prediction
can then be compared with the estimated number of FIRST quasars without SDSS
counterparts. This method essentially compares the $R_i$ distribution determined
for SDSS-FIRST quasars with the best guess of what the $R_i$ distribution may
be for FIRST-only quasars.

The main uncertainty in this method comes from the unknown fraction of quasars in
the FIRST-only sample; assuming the same fraction as determined for SDSS-FIRST
quasars (17\%), we find that there are 3.3 times as many quasars with \i$>$21.5 and
\t$<$16, than quasars with \i$<$21.5 and \t$<$16. On the other hand, with the extreme
assumption that {\em all} FIRST-only sources are quasars, the upper limit on this
ratio is 17. If the predicted ratio is much smaller than these estimates, then
a population of optically obscured quasars can be invoked to explain the discrepancy.

We compute the expected number of FIRST-only quasars from
\eq{
       N_{RL}(\i>21.5,t<16) = \int_{21.5}^\infty n(\i) f_{RL}(\i) \, d\i,
}
where $n(\i)$ describes the differential optical counts, and $f_{RL}(\i)$ is the
fraction of all quasars with optical magnitude \i\ that have \t$<$ 16. The latter
is obtained from (valid for \i$>$18.5)
\eq{
    f_{RL}(\i) = C_{RL}\, \left( {\int_{0.4(\i-16)}^\infty \phi(R_i) \, dR_i \over
                                \int_1^\infty \phi(R_i) \, dR_i} \right).
}
All quasars with \t$<$16 and \i$>$ 18.5 are radio-loud ($R_i>1$), hence the
index ``RL". Note that the \i\ integration only formally goes to infinity; the effective
range is \i $\la$ 26 because most quasars have $R_i \la 4$.

The constant $C_{RL}=0.08$, representing the fraction of quasars with $R_i>1$, and
the $R_i$ distribution, $\phi(R_i)$, are determined at the bright end.
The dashed line in the bottom panel in Figure \ref{quasartvsi} shows a best
Gaussian fit to the observed $\phi(R_i)$ distribution; it is centered at $R_i=2.8$
with $\sigma=0.8$. Deep optical quasar counts show flattening for \i \about20
(assuming \i \about \r), and follow a log($n$) = C + 0.3 \i\ relation in the range
20 $\la$ \i $\la$ 22.5 (Pei 1995). Pei's model, which explains well the available
observations for \i$<$22.5, predicts that this relation extends to at least
\i\about24. Based on this relation, we investigate two possibilities: a) this
relation is valid for \i $>$ 20, and b) it applies to 20 $<$ \i $<$22.5, and the
quasar counts drop to zero for \i $>$ 22.5. The second assumption is an extreme
relation that maximizes the number of quasars that can be attributed to the
``obscured" population.

This simple model predicts that the number of quasars with \t$<$16 and \i$>$ 21.5
is 45 times as large as the number of quasars with \t$<$16 and \i$<$ 21.5 for
possibility a), and 6 times as large for possibility b). These values are in
the same range as those implied by the counts of FIRST-only sources, and demonstrate
that there is no compelling need to invoke a significant population of anomalously
optically faint quasars. However, we caution that the uncertainty of the predicted counts
is large because of the unknown fraction of quasars in the FIRST-only sample, and
because quasar counts for 21.5 $<$ \i $<$ 26 are only weakly constrained. In particular,
it cannot be ruled out at a high confidence level that FIRST-only sources include a
substantial population of heavily obscured quasars (possibly as large as the population
selected by UV excess).

\vskip 0.2in

\appendix{{\bf Appendix C:  A Comment on the Definition of Radio Loudness}}

\vskip 0.1in

Two definitions of radio loudness are found in the literature. Schmidt (1970) proposed
the use of radio-to-optical flux ratios (or, equivalently spectral indices, see
Section 2.2) because it appeared that they are distributed independently of optical
luminosity and redshift. On the other hand, Peacock, Miller \& Longair (1986) proposed
that the radio luminosity should be used to quantify the radio-loudness.
The choice of a radio-loudness measure has important physical implications: the
radio-to-optical flux ratio is a proper quantity for analysis if the radio and
optical emissions are correlated; on the other hand, if the radio emission is
independent of optical emission, then the radio luminosity is the quantity of
interest. Here we compare both definitions for objects with redshifts from the SDSS
spectra, and find that they produce similar radio-loud/radio-quiet classifications. 
This similarity is a consequence of strong selection effects in flux-limited samples.

Figure \ref{quasarRvsM} displays the
absolute radio magnitude, $M_t$, vs. the radio-to-optical flux ratio, $R_i$, for the
1,154 quasars detected by FIRST. We compute $R_i$ using eq.~\ref{Rmt}, and $M_t$ from
\eq{
          M_t = -2.5 \log\left({L_{radio} \over L_{AB}}\right),
}
where
\eq{
          L_{radio} = {4 \pi \, D_L^2 \over (1+z)^{1+\alpha_r}} \, F_{int},
}
is the specific radio luminosity, and $L_{AB} = 4.345\times10^{13}$ W Hz$^{-1}$
is the specific luminosity of a source whose specific flux is 3631 Jy at a distance
of 10 pc. We assumed a radio spectral index $\alpha_r = -0.5$, and for
consistency with Stern {\em et al.} (2000) adopted $H_o$ = 50 km s$^{-1}$
Mpc$^{-1}$, and a $\Omega_M=1$, $\Omega_\Lambda=0$ universe in computing the
luminosity distance, $D_L$. As evident, $R_i$ and $M_t$ are well correlated.
Furthermore, the separations between the radio-loud and radio-quiet objects
suggested by other workers ($R_i$ \about 1 and $L_{radio} \about 10^{24}$
h$_{50}^{-2}$ W Hz$^{-1}$, corresponding to $M_t=-25.9$) appear qualitatively
consistent with each other (the two classifications are the same for 88\% of the
objects).

It is somewhat surprising that a dimensionless quantity that measures the shape of
SED, $R$, is so well correlated with the (dimensional) radio power, $L_{radio}$.
Furthermore, $R$ is practically independent of redshift, while $L_{radio}$ is a
strong function of redshift at constant apparent magnitude. In principle, such a
correlation could be a consequence of a physical relationship between optical and
radio emission; for example, the luminosity of main-sequence stars is correlated
with the shape of the optical SED. However, we find instead that this correlation is
due to several selection effects caused by the quasar redshift distribution and
very different slopes of the optical and radio number counts (0.87 vs. 0.14,
respectively), as follows.

For a sample following a number count relation $\log(n) = C + k\,m$, 90\% of the sample
is within $k^{-1}$ mag from the faint cutoff. Thus, 90\% of the SDSS-FIRST sample
is within \about1 mag of the optical faint cutoff, and within \about7 mag of the
radio faint cutoff. The minimum value of $R_i$ is $0.4\,(i_{min}-t_{max})$, where
$t_{max}$\about16 is the radio faint limit, and $i_{min}$ is the optical bright limit.
Since 90\% of the sources are within \about1 mag of the faint optical limit (\about 19),
only 10\% of the sample can have $R_i \la 0.8$, as observed. The maximum value of
$R_i$ is $0.4\,(i_{max}-t_{min})$, where $i_{max}$\about19 is the optical faint limit,
and $t_{min}$ is the radio bright limit. A negligible number of sources are brighter
than $t=9$, and thus practically no sources have $R_i > 4$.

Given $R_i$, the variance in $M_t$ (the vertical width of the displayed
correlation; $\sigma$ \about1.5 mag) reflects the variances in \i\ and the
distance modulus ($M_t = \i - DM - 2.5\,R_i$). The latter dominates the scatter in
$M_t$ because the
scatter in \i\ is small due to the steep optical counts. The scatter in the distance
modulus is determined by the distribution of redshifts; the distribution of
distance moduli has a median of 44.5 mag, and an equivalent Gaussian width
of 1.4 mag. The upper limit is a
consequence of the upper limit on redshift, and the lower limit is due to the fast
increase of the number of quasars per unit redshift interval with redshift.
This narrow distribution of distance moduli explains the observed upper (bright) and
lower (faint) limits on $M_t$ for a given $R_i$. The slope of the observed correlation,
$dM_t/dR_i = -2.5$, simply reflects the relationship between $R_i$ and $t$.
We conclude that the difference in the slopes of the radio and
optical number count relations, together with the observed redshift
distribution, is responsible for the apparent correlation between $M_t$ and $R_i$,
and there is no intrinsic correlation.

Due to this selection effect, the observed bimodal distribution of $R$ maps into a
{\em biased} bimodal distribution of $L_{radio}$. An unbiased distribution of $L_{radio}$
cannot be determined with the sample discussed here. In particular, a distribution
of data points in $L_{radio}$ vs. redshift plane (analogous to Figure \ref{Mredshiftquasar})
has a strong bias in $R$ that depends on both quantities (only sources with
$R_i < 0.4\,[i^*_{cutoff}-t(M_t,z)]$ are sampled, where $i^*_{cutoff}$ is the
optical faint flux limit). For the same reason, the radio luminosity distributions for
galaxies (Section \ref{ORgalaxies}) and quasars from SDSS-FIRST sample cannot be directly
compared due to significantly different redshift ranges over which they are sampled.
A proper determination of the radio luminosity function requires identification and redshift
determination for {\em all} FIRST sources brighter than some radio flux limit, as discussed
by e.g. Willot {\em et al.} (1998) and references therein.

\newpage

\begin{figure}
\caption{The SDSS color-color
and color-magnitude diagrams for $\about$ 300,000 objects observed in 50 deg$^2$
of sky. The top two panels show color-color diagrams for objects with photometric
errors less than 0.1 mag in the plotted bands. The unresolved sources are shown as
dots, and the distribution of resolved sources is shown by linearly spaced density
contours. The low-redshift quasars ($z$ $\la$ 2.5), selected by their blue \ug\
colors, are shown as circles. Almost all of the unresolved sources marked as dots are
stars (a small fraction may be quasars and compact galaxies), with approximate spectral 
types as marked. The bottom two panels show color-magnitude diagrams for unresolved 
(left) and resolved (right) sources.\label{SDSS3}}
\end{figure}

\begin{figure}
\caption{The \r\ vs. \gr\ color-magnitude diagrams for \about190,000 galaxies
separated by their \ur\ color: the upper panel shows \about 126,000 blue galaxies
with \ur $<$ 2.22, and the lower panel shows \about64,000 red galaxies with \ur $>$ 2.22.
The faint end distribution shape is due to magnitude cutoffs. The dashed lines outline
several characteristic regions which are useful when analyzing the properties of
galaxies (see text for details). \label{GALcmd}}
\end{figure}

\begin{figure}
\caption{The top panel shows
the regions included in the SDSS imaging data studied here (1230 deg$^2$). A smaller
region (774 deg$^2$) with SDSS spectroscopic data is shown in the bottom panel. The
regions are outlined by sparse sampling the source positions. \label{footprint}}
\end{figure}

\begin{figure}
\caption{The top panel shows
the size measure, $\log(\theta^2)$, vs. the radio AB magnitude, $t$, for 28,476 FIRST
sources from a 325 deg$^2$ large region of sky (EDR sample). The diagonal cutoff
running from the top to the lower right corner is due to the FIRST faint limit. The
bottom panel shows differential $t$ distributions (``counts") for all sources
(circles), and for the 9823 sources with $\log(\theta^2) > 0.1$ (triangles). The
dashed line is a best linear fit to the counts of all sources in the $11.5 < t <
15.5$ range (see text).  \label{rawFIRST}}
\end{figure}

\begin{figure}
\caption{The top panel shows the
distributions of the distance between the SDSS and FIRST positions for the 10,084
close pairs from a 325 deg$^2$ region; the 1,999 optically unresolved source are 
marked by triangles, and the 8,085 resolved sources by circles. The vertical dashed 
line shows the adopted cutoff (1.5 arcsec) for positional association that results in an
85\% complete sample with a contamination of 3\%. The middle and bottom
panels show differences in equatorial coordinates for sources with \r$<$20 and
\t$<$15. The declination shows an offset of 0.12 arcsec (see text).
\label{offset}}
\end{figure}

\begin{figure}
\caption{The top panel compares
the differential counts of FIRST sources with an SDSS identification (triangles)
to those without (dots), as a function of radio AB magnitude, \t. The two
lines show best linear fits to the counts in the 11.5 $< t < $ 15.5 range. The
bottom panel compares the distributions of $\theta$, which is a rough measure of
the source radio size, for sources with $t < 15$ (same notation as in top panel).
\label{m-unmR} }
\end{figure}

\begin{figure}
\caption{The top panel shows the \r\ vs. \rz\ color-magnitude diagram for the
optically identified FIRST sources. The 23,898 resolved sources are shown as contours
and 5,623 unresolved sources as dots. The bottom panel shows the \rz\ distributions for
sources with 21$<$\r$<$ 21.5; the triangles correspond to 625 unresolved sources,
and the circles to 2,437 resolved sources. Optically resolved sources tend to be red,
while unresolved sources tend to be blue, even close to the faint limit.
\label{colorSG} }
\end{figure}

\begin{figure}
\caption{Comparison of
the radio properties for SDSS-detected quasars and galaxies. The top panel
compares the differential counts as a function of the radio magnitude for 3,142
quasars, marked by dots, and 16,109 galaxies, marked by triangles. The dashed lines
show the best fits discussed in text. The sum of counts for quasars and galaxies,
multiplied by 5.6 to account for the matching fraction, is shown as open squares,
and compared to the counts of all FIRST sources, shown by the solid line. Note the
similarity between the two distributions. The bottom panel displays the
distributions of the radio size parameter $\theta$ for quasars (triangles) and galaxies
(circles) with $t < 15$.
\label{radioCounts} }
\end{figure}

\clearpage


\begin{figure}
\caption{The \u\ vs. \ug\ color-magnitude diagram and three color-color diagrams
for stars, shown by contours, and for optically unresolved radio sources brighter
than \r=20.5, shown as symbols. Open circles mark
the 383 radio sources with UV excess satisfying \u $<$ 21 and \ug $<$ 0.7. Open squares
mark the 86 radio sources with non-stellar colors but without the UV excess, and the
solid squares mark the 68 radio sources with colors indistinguishable from stellar.
The dashed lines outline the boundaries of the 4-dimensional stellar locus used
to select sources with stellar colors.
\label{quasarcc}
}
\end{figure}

\begin{figure}
\caption{Examples of SDSS
spectra (spectral resolution \about 2000) for optically unresolved sources with
stellar colors and FIRST detections within 1.5 arcsec. The object's name, redshift
estimate, and classification is marked in each panel. \label{weirdQSO} }
\end{figure}

\begin{figure}
\caption{The top panel shows the dependence of quasar \gi\ color
on redshift. The distribution of 6,567 optically selected quasars with
\i $<$ 18.5 is shown by contours; those that are resolved (2,095) are marked
by crosses. The 280 FIRST-detected quasars with $R_i>1$ (radio-loud) are
shown as filled circles, and the 161 FIRST-detected quasars with $R_i<1$
(radio-quiet) are shown as open circles.
The thick solid line shows the median \gi\ color of all optically
selected quasars in the redshift range 1--2, which is subtracted from the \gi\
color to obtain a color excess. The bottom panel shows the distribution of
the \gi\  color excess for 2,265 quasars in that redshift range by solid squares
(without error bars), and for 102 radio-loud quasars by circles.
\label{quasargi}
}
\end{figure}

\begin{figure}
\caption{
Optical spectral indices  $\alpha$ ($F_\nu \propto \nu^\alpha$) determined from
SDSS spectra for 6,868 quasars that are brighter than \i=19 (filled circles). The
$\alpha$ distribution for a subsample of 440 radio-loud quasars (triangles with
error bars), is skewed towards more negative values (redder optical spectra).
\label{alphaD}
}
\end{figure}

\begin{figure}
\caption{The open squares
show differential counts for optically unresolved and spectroscopically
confirmed SDSS quasars. The turnover at \i \about19 is a selection effect due to a
flux limit for spectroscopic targeting, as indicated by the open triangles that
show the counts of UVX-selected optically unresolved SDSS sources brighter than
\u=21 (displayed only for \i$>$17.5 for clarity). The counts for 1154
FIRST-detected optically unresolved objects are shown by filled circles. The solid
triangles show counts for a subset of 969 FIRST-detected radio-loud sources with
$R_i > 1$. The dashed lines are the best linear fits in the 15.5 $< \i < $ 18.0
range described in the text. For \i$<$18.5, the fraction of FIRST-detected quasars
is 13\%, and the fraction of radio-loud quasars is \about8\%.
\label{quasaroptcounts} }
\end{figure}

\begin{figure}
\caption{The top panel shows
the $t$ (radio magnitude) vs. \i\ (optical magnitude) distribution of the 3,066
optically unresolved SDSS sources detected by FIRST with \i$<$21 and $t<16.5$.
The diagonal solid line shows the traditional radio loud/quiet division
line ($R_i = 1.0$), and the four short-dashed lines show $R_i$ = 0, 2, 3, and 4,
as marked. In the lower panel, the histogram marked by open squares shows the $R_i$
distribution for 359 sources with \i $<$ 18. The two histograms marked by filled circles
and triangles show the $R_i$ distribution for the 670 sources selected from
the two strips defined by the three dot-dashed lines shown in the top panel.
The dashed line is a best Gaussian fit to the sum of these two histograms
for $R_i>1$.
\label{quasartvsi} }
\end{figure}

\begin{figure}
\caption{The upper panel
shows the redshift and
absolute magnitude distribution ($M_i$) for 4,472 optically selected quasars with
\i$<$18.5, marked by dots, and for a subsample of 280 radio-loud quasars ($R_i >$ 1),
marked by squares. The middle panel shows the distribution of absolute
magnitudes for all quasars (dots without error bars), and for the radio
subsample (squares) from the same redshift range as shown in the top panel. 
The bottom panel shows the corresponding redshift distribution (same notation). 
There is no significant difference between the distributions for the radio subsample 
and for the whole sample. \label{Mredshiftquasar}}
\end{figure}



\begin{figure}
\caption{The \r\ vs. \gr\ color-magnitude diagram for SDSS-FIRST galaxies,
shown by dots, compared to the distribution of all SDSS galaxies, shown by
linearly-spaced contours. The top panel shows 19,496 SDSS-FIRST galaxies
brighter than \r=21.5, and the bottom panel shows 4,300 galaxies for
which SDSS spectra are available (the area covered by the latter subsample
is 63\% of the area covered by the former). The dashed lines outline regions
with different galaxy morphology and fraction of radio galaxies, as listed
in Table 1.
\label{gal-colmag-gr}
}
\end{figure}

\begin{figure}
\caption{Mosaic of true-color g-r-i color composite images for a subsample of SDSS-FIRST
galaxies with \r $<$ 17.5 and \ur $<$ 2.22. The image size is 1x1 arcmin, with East on top
and North towards right. The position of the associated FIRST source is marked by a cross. Available as gif file.
\label{colormosaic} }
\end{figure}

\begin{figure}
\caption{The SDSS $r$-band images
for the same galaxies as in previous Figure. The stretch is chosen to emphasize
galactic nuclei, and the display is negative. Available as gif file.\label{rmosaic} }
\end{figure}

\clearpage

\begin{figure}
\caption{The optical
color-color diagrams for 4,152 SDSS-FIRST galaxies with \r$<$17.5, shown as dots,
compared to the distribution of all SDSS galaxies, shown by linearly-spaced
contours. The bottom right panel shows the correlation between the redshift and
the \ri\ color for galaxies with spectra (141,920 for the full sample and
4300 for radio-galaxies). \label{GALccd}}
\end{figure}

\begin{figure}
\caption{The top left panel
shows the distribution of the 91,422 galaxies (contours) with \r $<$ 17.5 (contours)
in the \r\ vs. \ur\ color-magnitude diagram; a subsample of 2,563 radio galaxies
by dots. The top right panel shows the arginal \ur\ distributions for all (dashed line)
and radio (symbols and solid line) galaxies; the latter are redder. The bottom left
panel compares the distributions of radio galaxies and all galaxies in the \ur\
color-redshift plane. The bottom right panel compares the redshift distributions of
for two color-selected subsamples: dashed line (all) and open circles (radio) for
galaxies with \ur $<$ 2.22, and solid line (all) and solid squares (radio) for galaxies
with \ur $>$ 2.22. Radio galaxies in a flux limited sample are biased towards larger
redshifts. \label{GALur} }
\end{figure}

\begin{figure}
\caption{The top left panel shows the \r\ vs. \ur\ color-magnitude distribution
for galaxies with \r $<$ 17.5, redshift in the range 0.08--0.12, and, for radio
galaxies, $R_r > 0.4$; 219 radio galaxies are shown by dots and all 31,226 galaxies by contours.
The top right panel compares the \ur\ histograms for radio (symbols) and all
(dashed line) galaxies. The middle and bottom left panels compare the apparent and
absolute magnitude distributions for blue (\ur$<$2.22) galaxies, and the right panels
for red (\ur$>$2.22) galaxies. All galaxies are marked by thin solid line, and radio
galaxies by symbols. Radio galaxies have a different luminosity distribution from other
galaxies.
\label{optlum}}
\end{figure}

\begin{figure}
\caption{The distribution of
SDSS-FIRST galaxies with \r$<$17.5 and $t<16.0$, from two redshift slices
(0.03--0.07 in the left column and 0.08--0.12 in the right column), in the $t$ vs.
\r\ diagrams. The top panels show galaxies with \ur$<$ 2.22, and the middle
panels show galaxies with \ur$>$ 2.22. The three dashed lines in the top four
panels show constant radio-to-optical flux ratios of 0, 1, and 2, as marked. The
bottom panels show the $R_r$ distributions for objects from the strip defined by
the two dot-dashed lines in each panel; only $R_r > 0$ points are complete.
\label{GALtr}}
\end{figure}

\begin{figure}
\caption{The distribution of
SDSS-FIRST galaxies with $t<16.0$, from two redshift slices (0.28--0.32 in the
top panel and 0.38--0.42 in the middle panel), in the $t$ vs. \z\ diagrams. The
three dashed lines in the top two panels show constant radio-to-optical flux
ratios of 0, 1, and 2, as marked. The bottom panel shows the $R_z$ distributions
for objects from the strip defined by the two dot-dashed lines in the top panel
(circles: lower redshift, triangles: larger redshift).
\label{GALtrX}}
\end{figure}


\begin{figure}
\caption{Examples of SDSS spectra for radio galaxies discussed here. The galaxy
name, redshift, color and brightness is marked in each panel.
\label{spectra1} }
\end{figure}

\begin{figure}
\caption{Additional examples of SDSS spectra for radio galaxies discussed here.
The galaxy name, redshift, color and brightness is marked in each panel.
\label{spectra2} }
\end{figure}

\begin{figure}
\caption{The line diagnostic
diagrams separating AGNs and starburst galaxies. The distributions of
16,325 SDSS galaxies with \r$<$17.5 are shown by linearly spaced contours. The
650 SDSS-FIRST galaxies selected by the same criteria are shown as circles. The dashed
lines, obtained theoretically by Kewley {\em et al.} (2001), and the dot-dashed
line in the top panel, separate AGNs from starburst galaxies. There are \about18
starburst galaxies for each AGN in the entire SDSS sample; for the SDSS-FIRST sample,
this ratio is \about2.4. \label{AGNs} }
\end{figure}

\begin{figure}
\caption{The comparison
of the magnitude, color and redshift distributions for SDSS-FIRST galaxies
classified as AGN (143, triangles) and starburst galaxies (422, dots) using the
line diagnostic diagrams. The distribution of the control sample of SDSS
galaxies is shown by contours in the top panel, and by the short-dashed
(starburst galaxies) and long-dashed (AGNs) lines in the other two panels.
AGNs are redder than starbursts by \about 0.6 mag for both radio sample
and the control sample.
\label{AGNrgr} }
\end{figure}

\begin{figure}
\caption{Comparison of the radio properties for SDSS-FIRST galaxies
classified as AGNs (triangles) and starburst galaxies (circles) using emission
line strengths. The top panel compares the differential counts as a function of
radio magnitude, and the bottom panel displays the distributions of the
radio concentration parameter $\theta$ for sources with $t < 15$. The cumulative
counts for $t < 15$ are similar, while there are three times more starburst
galaxies than AGNs in the $15 < t < 16$ range. Starburst galaxies tend to have
larger radio concentration parameters than do AGNs.
\label{radioCountsAGNvsSF}}
\end{figure}

\begin{figure}
\caption{The top two panels
show the \iz\ vs. \ri\ color-color diagrams for SDSS sources from \about100 deg$^2$
of sky, and selected from two magnitude bins: \r$<$17.5 (left, \about20,000 sources) and
21.4 $<$ \r $<$ 21.5 (right, \about20,000 sources). The distribution of unresolved sources
is shown as linearly-spaced contours, and resolved sources are shown as dots. The bottom
two panels show the distribution of \about22,000 SDSS-FIRST sources from 1230 deg$^2$ of
sky in the \r\ vs. \ri\ and \iz\ vs. \ri\ diagrams (same notation).
\label{highz}}
\end{figure}

\begin{figure}
\caption{The distribution of
SDSS-FIRST quasars in the radio-luminosity (absolute magnitude $M_t$) vs.
radio-to-optical flux ratio, $R_i$, plane. All 1154 SDSS-quasars are marked by open
circles. A subsample of 531 objects with \i$<$19 and $t<$ 15.5 are marked by solid
circles. The vertical and horizontal lines are the traditional division lines
between the radio-loud and radio-quiet quasars.  Note that most quasars have the same
classification in both $M_t$ and $R_i$ based schemes.
\label{quasarRvsM} }
\end{figure}

\end{document}